\documentclass[twocolumn]{aastex61}

\begin{document}


\title{Optical, near-IR and sub-mm IFU Observations of the nearby dual AGN Mrk\,463}

\author[0000-0001-7568-6412]{Ezequiel Treister}
\affiliation{Instituto de Astrof\'{\i}sica  and Centro de Astroingenier{\'{\i}}a, Facultad de F\'{i}sica, Pontificia Universidad Cat\'{o}lica de Chile, Casilla 306, Santiago 22, Chile}

\author{George C. Privon}
\affiliation{Instituto de Astrof\'{\i}sica  and Centro de Astroingenier{\'{\i}}a, Facultad de F\'{i}sica, Pontificia Universidad Cat\'{o}lica de Chile, Casilla 306, Santiago 22, Chile}
\affiliation{Department of Astronomy, University of Florida, 211 Bryant Space Sciences Center, Gainesville, 32611 FL, USA}

\author{ Lia F. Sartori}
\affiliation{Institute for Astronomy, ETH Z\"urich, Wolfgang-Pauli-Stra\ss e 27, CH-8093 Zurich, Switzerland}

\author{Neil Nagar}
\affiliation{Universidad de Concepci\'{o}n, Departamento de Astronom\'{\i}a, Casilla 160-C, Concepci\'{o}n, Chile}

\author{Franz E. Bauer}
\affiliation{Instituto de Astrof\'{\i}sica  and Centro de Astroingenier{\'{\i}}a, Facultad de F\'{i}sica, Pontificia Universidad Cat\'{o}lica de Chile, Casilla 306, Santiago 22, Chile}
\affiliation{Millennium Institute of Astrophysics (MAS), Nuncio Monse{\~{n}}or S{\'{o}}tero Sanz 100, Providencia, Santiago, Chile}
\affiliation{Space Science Institute, 4750 Walnut Street, Suite 205, Boulder, Colorado 80301}

\author{ Kevin Schawinski}
\affiliation{Institute for Astronomy, ETH Z\"urich, Wolfgang-Pauli-Stra\ss e 27, CH-8093 Zurich, Switzerland}

\author{Hugo Messias}
\affiliation{Joint ALMA Observatory, Alonso de C\'ordova 3107, Vitacura 763-0355, Santiago, Chile}
\affiliation{European Southern Observatory, Alonso de C\'ordova 3107, Vitacura, Casilla 19001, 19 Santiago, Chile}

\author{Claudio Ricci}
\affiliation{Instituto de Astrof\'{\i}sica  and Centro de Astroingenier{\'{\i}}a, Facultad de F\'{i}sica, Pontificia Universidad Cat\'{o}lica de Chile, Casilla 306, Santiago 22, Chile}
\affiliation{Kavli Institute for Astronomy and Astrophysics, Peking University, Beijing 100871, China}
\affiliation{Chinese Academy of Sciences South America Center for Astronomy and China-Chile Joint Center for Astronomy, Camino El Observatorio 1515, Las Condes, Santiago, Chile}

\author[0000-0002-1912-0024]{Vivian U}
\affiliation{University of California Chancellor's Postdoctoral Fellow. Department of Physics and Astronomy, University of California, Riverside, 900 University Avenue, Riverside, CA 92521, USA}

\author[0000-0002-0930-6466]{Caitlin Casey}
\affiliation{Department of Astronomy, The University of Texas at Austin, 2515 Speedway Boulevard Stop C1400, Austin, TX 78712, USA}

\author{Julia M. Comerford}
\affiliation{Department of Astrophysical and Planetary Sciences, University of Colorado Boulder, Boulder, CO 80309, USA}

\author{Francisco Muller-Sanchez}
\affiliation{Department of Astrophysical and Planetary Sciences, University of Colorado Boulder, Boulder, CO 80309, USA}

\author{Aaron S. Evans}
\affiliation{National Radio Astronomy Observatory, 520 Edgemont Road, Charlottesville, VA 22903, USA}
\affiliation{Department of Astronomy, 530 McCormick Road, University of Virginia, Charlottesville, VA 22904, USA}

\author{Carolina Finlez}
\affiliation{Universidad de Concepci\'{o}n, Departamento de Astronom\'{\i}a, Casilla 160-C, Concepci\'{o}n, Chile}

\author{Michael Koss}
\affiliation{Institute for Astronomy, ETH Z\"urich, Wolfgang-Pauli-Stra\ss e 27, CH-8093 Zurich, Switzerland}
\affiliation{Eureka Scientific Inc., 2452 Delmer St. Suite 100, Oakland, CA 94602, USA}

\author{David B. Sanders}
\affiliation{Institute for Astronomy, 2680 Woodlawn Drive, University of Hawaii, Honolulu, HI 96822, USA}

\author[0000-0002-0745-9792]{C. Megan Urry}
\affiliation{Yale Center for Astronomy \& Astrophysics, Physics Department, New Haven, CT 06520, USA}

\correspondingauthor{Ezequiel Treister}
\email{etreiste@astro.puc.cl}

\begin{abstract}
We present optical and near-IR Integral Field Unit (IFU) and ALMA band 6 observations of the nearby dual Active Galactic Nuclei (AGN) Mrk\,463.  At a distance
of 210 Mpc, and a nuclear separation of $\sim$4 kpc, Mrk\,463 is an excellent laboratory to study the gas dynamics, star formation processes and 
supermassive black hole (SMBH) accretion in a late-stage gas-rich major galaxy merger. The IFU observations reveal a complex morphology, 
including tidal tails, star-forming clumps, and emission line regions. The optical data, which map the full extent of the merger, show evidence for a biconical outflow and 
material outflowing at $>$600 km s$^{-1}$, both associated with the Mrk\,463E  nucleus, together with large scale gradients likely related to the ongoing galaxy merger. 
We further find an emission line region $\sim$11 kpc south of Mrk\,463E that is consistent with being photoionized by an AGN. Compared to the current AGN 
luminosity, the energy budget of the cloud implies a luminosity drop in  Mrk\,463E by a factor 3--20 over the last 40,000 years. The ALMA observations of $^{12}$CO(2-1) 
and adjacent 1\,mm continuum reveal the presence of $\sim$10$^{9}$M$_\sun$ in molecular gas in the system. The molecular gas shows velocity gradients of 
$\sim$800 km/s and $\sim$400 km/s around the Mrk\,463E and 463W nuclei, respectively. We conclude that in this system the infall of $\sim$100s $M_\odot$/yr
of molecular gas is in rough balance with the removal of ionized gas by a biconical outflow being fueled by a relatively small, $<$0.01\% of accretion onto each
SMBH. 
\end{abstract}

\keywords{galaxies: interactions, active, Seyfert, indvidual (Mrk 463)}

\section{Introduction} 
\label{sec:intro}

For the last $\sim$30 years there has been growing evidence for a strong connection between major, $<$3:1 mass ratio, galaxy mergers and simultaneous episodes of 
strong star formation and significant central supermassive black hole (SMBH) growth \citep[e.g.,][]{sanders88,barnes91,di-matteo05}. Early observations of nearby 
galaxies carried out with the {\it Infrared Astronomical Satellite} ({\it IRAS}) have shown that the most luminous IR sources appear to be associated with interacting  
galaxies, as evidenced by the presence of tidal tails, disrupted morphologies and other features \citep{soifer84}. Theoretical models and computational simulations 
\citep[e.g.,][]{barnes91,mihos96,springel05} have shown that a major galaxy merger provides a very efficient mechanism to drive gas to the nuclear regions 
of the resulting system, which can in turn fuel the observed star formation bursts and SMBH growth. As a consequence, it is natural to expect that major 
mergers can play a fundamental role in galaxy evolution \citep[e.g.,][]{hopkins06}. 

Major galaxy mergers can also explain the observed correlations between the mass of the central SMBH and physical properties of its host galaxy 
\citep[e.g.,][]{di-matteo05,di-matteo08}. However, a key element in setting up this physical connection are so-called feedback effects. Such effects are regularly 
invoked by semi-analytical models in order to reproduce even basic galaxy properties, such as the luminosity function 
\citep[e.g.,][and many others since]{kauffmann98}, as they provide additional sources of energy required to prevent runaway star formation episodes 
\citep{benson03,croton06,schawinski06}. In particular, nuclear activity can play a critical role in the regulation of star formation, as observed in some nearby 
galaxies \citep[e.g.,][]{alatalo15}.

\citet{treister12} found that while most AGN activity is triggered by internal, secular processes and minor galaxy mergers, major mergers are directly linked to the 
most luminous AGN, quasars, and therefore are responsible for most ($\sim$60\%) of the SMBH accretion across the cosmic history. A natural consequence 
of this scenario is that dual AGN --- i.e., systems in which the two nuclear SMBHs are growing simultaneously at separations of $<$10 kpc --- should be relatively 
common \citep[e.g.,][]{volonteri03,fu11a,van-wassenhove12}. This particular stage in a major galaxy merger, albeit short --- $\sim$hundreds Myears --- 
\citep{van-wassenhove12,blecha13}, is very relevant for galaxy evolution. As shown by \citet{koss12}, both the fraction of dual AGN and their individual X-ray 
luminosities peak at nuclear separations $<$10~kpc. This implies that significant and rapid SMBH growth can be directly associated with the dual AGN phase, which 
in turn might be associated with strong feedback effects on the interstellar medium \citep[e.g.,][]{rupke11,veilleux13}. 

Furthermore, the relative frequency of dual systems can be used to constrain the AGN duty cycle in merger-triggered events. In a rather extreme scenario, if the 
lifetime of the AGN/quasar phase is similar to the merger timescale, we should expect every merger-triggered system to host a dual AGN. However, the observed 
fraction of dual AGN is significantly lower, $\sim$2\% \citep{liu11,shen11}. This can be explained at least in part by the difficulty in detecting and confirming 
observationally these dual AGN. Confirming the dual AGN nature even for nearby sources often requires high spatial resolution radio, near-IR and/or X-ray 
observations \citep[e.g.][]{fu11,koss12,muller-sanchez16,mcgurk15}. Furthermore, in many cases the nuclear regions in these systems are subject to large amounts 
of obscuration \citep{hopkins05,scoville17,ricci17}, making their detections even more challenging. Nonetheless, many merger systems lack dual nuclei despite 
adequate data. This suggests the AGN duty cycle is shorter than the merger time scale.

In order to study the physical properties of the ionized, atomic and molecular gas and the dust in confirmed dual AGN in the local Universe, $z$$<0.1$, we started 
the Multiwavelength Observations of Dual AGN (MODA) program\footnote{\url{http://moda.astro.puc.cl/}}. The aim of MODA is to study the growth and co-evolution of 
galaxies and their SMBHs during merger events, particularly at the later merger stages when feedback and nuclear activity likely peak. MODA currently focuses on 
17 confirmed dual AGN distributed across the sky. Our program combines:  ({\it i}) Atacama Large Millimeter Array (ALMA) observations of molecular gas tracers and 
dust continuum mm/sub-mm emission, ({\it ii}) Very Large Telescope (VLT)/ Spectrograph for INtegral Field Observations in the Near Infrared (SINFONI) and 
Keck/OSIRIS (OH- Suppressing Infra-Red Imaging Spectrograph) maps of $H_2$, Pa$\alpha$, Br$\gamma$, [Si VI], and other near-IR emission 
lines, and ({\it iii}) VLT Multi Unit Spectroscopic Explorer (MUSE) optical Integral Field Unit (IFU) spectroscopic observations, aimed primarily to map the 
atomic emission lines of H$_\alpha$, $H_\beta$, [O{\sc iii}] and other atomic species, together with Na D in absorption, in order to measure gas and stellar 
kinematics, and constrain stellar populations and star formation histories. In addition, we are now in the process of adding mid-IR observations from the VLT 
spectrometer and imager for the mid-infrared (VISIR) and SOFIA (Stratospheric Observatory for Infrared Astronomy) to study the properties of the hot, $\sim$300-1,000K, 
dust in these systems, compute accurate nuclear mid-IR to X-ray flux ratios and obtain accurate estimates of the bolometric luminosities of each AGN.

Here, we focus on one of the sources in the MODA sample, Mrk\,463, for which we recently finished obtaining multi-wavelength IFU observations. 
Mrk\,463 is a nearby, $z$=0.0504 \citep{falco99} or $d$$\simeq$210Mpc, Ultra-Luminous Infrared Galaxy (ULIRG; \citealp{sanders96}). It was classified as a dual 
AGN based on {\it Chandra} X-ray observations reported by \citet{bianchi08}. The two nuclei are separated by 3.8$\pm$0.01$''$, which roughly corresponds to 
3.8 kpc. The {\it Chandra} observations reveal that both nuclei, Mrk\,463E and Mrk\,463W, respectively, have relatively low X-ray luminosities of 
$L_{\rm X}$=1.5$\times$10$^{43}$~erg/s and 3.8$\times$10$^{42}$~erg/s in the 2-10 keV band and are heavily obscured with 
$N_{\rm H}$=7.1$\times$10$^{23}$~cm$^{-2}$ and 3.2$\times$10$^{23}$~cm$^{-2}$. The AGN nature of the eastern nucleus was independently established
by Spitzer spectroscopic observations, as presented by \citet{farrah07}. Given its distance, nuclear separation, infrared luminosity and existing 
multi-wavelength observations, Mrk\,463 joins NGC6240 \citep{komossa03}, Mrk 739 \citep{koss11a} and Mrk 266 \citep{mazzarella12} as prime targets 
for understanding how SMBHs gain mass and how black hole growth can impact their host galaxies in the dual AGN phase of a major galaxy merger.

In this paper we present optical and near-IR IFU observations of the nearby dual AGN Mrk\,463, obtained with the VLT MUSE and SINFONI, and ALMA Cycle 2 
observations of $^{12}$CO(2-1) and adjacent sub-mm continuum in this system. The main goal is to study the morphology and kinematics of the gas in all its 
phases using atomic and molecular emission lines as tracers, in order to understand the behavior of the gas that is simultaneously feeding the SMBH and 
fueling star formation during a major galaxy merger. In \S2, we describe all the different datasets and data reduction strategies used. In \S3, we describe 
the multi-wavelength morphological properties of this system, while in  \S4 we focus on the kinematics. The discussion and conclusions are presented 
in \S5 and \S6, respectively. Throughout this paper, we assume a $\Lambda$CDM cosmology  with $h_0$=0.7, $\Omega_m$=0.27 
and $\Omega_\Lambda$=0.73 \citep{hinshaw09}.

\section{Data} 
\label{sec:data}

\subsection{MUSE}
\label{sec:muse}

MUSE is a state-of-the-art second generation VLT IFU instrument \citep{bacon10}. Its relatively large field of view, $\sim$1 arcmin$^2$, and broad wavelength 
coverage, $\sim$4800-9300\AA, are particularly well suited for observing the entire physical extents of low redshift systems (e.g., physical sizes of tens of kpc 
in a single pointing) across a large spectral range (simultaneous coverage of the $H_\beta$, $H_\alpha$ and [O{\sc iii}] lines, among others). The resolving power
of MUSE ranges from 1770 at 4800\AA~ to 3590 at 9300\AA, which corresponds to velocity resolutions of $\sim$80 to $\sim$170 km/s.

Mrk\,463 was observed by the VLT/MUSE as part of program 095.B-0482 (PI: E. Treister). Four other nearby dual AGN were also observed as part of this 
program and will be presented on subsequent papers. Mrk\,463 was observed on July 18, 2015 in two 58 min observation blocks (OBs), IDs 1182834 and 1182837, 
each comprised of  three on-target integrations of 976 s plus overheads. Sky conditions were clear, with $<$50\% lunar illumination, and a reported ambient seeing 
$<$0.8$''$.  The final combined, processed and calibrated VLT/MUSE data cube for Mrk\,463 has an effective exposure time of 5856 seconds ($\sim$1.6 hours), an 
average airmass of $\sim$1.5 and an average FWHM of $\sim$0.7$''$ based on measurements of stars in the field on the reconstructed white filter 
image.

Data reduction was carried out using the ESO MUSE pipeline version 1.6.2\footnote{Available at \url{https://www.eso.org/sci/software/pipelines/muse/}}
\citep{weilbacher14}, in the ESO {\it Reflex} graphical environment \citep{freudling13}. The data reduction stages are standard and include bias and 
dark current subtraction, flat fielding, wavelength and flux calibration, and astrometric correction. Sky background was subtracted using emission-free regions of 
the resulting data cube. The process is however complicated by the large data volume. The MUSE 1$'$$\times$1$'$ field of view has 300$\times$300 spatial bins (spaxels) and 4000 spectral bins, such that an individual MUSE frame is a few Gbytes in size and data reduction requires several Gbytes for intermediate products 
and multiple CPU cores to run. First-look visualizations of the resulting combined data cube were done using the 
QFitsview\footnote{\url{http://www.mpe.mpg.de/~ott/QFitsView/}} software, but not used in our subsequent work. Further analysis, including emission 
line fitting, velocity maps, etc. of well-isolated narrow emission lines was carried out using IDL {\it Fluxer}\footnote{\url{http://www.ciserlohe.de/fluxer/fluxer.html}} 
tool version 2.7. More complex line fits, including several overlapping components were carried out using the Pyspeckit Python package \citep{ginsburg11}, as 
Fluxer only allowed a maximum of two simultaneous gaussian fits.

\subsection{SINFONI}
\label{sec:sinfoni}

We obtained near-IR IFU observations of Mrk\,463 using the Spectrograph for INtegral Field Observations in the Near Infrared (SINFONI; \citealp{eisenhauer03}) camera mounted at the VLT, as part of program 093.B-0513 (PI: S. Cales). This program aimed to observe four confirmed $z$$<$0.1 dual AGN that are  visible from the southern hemisphere, one of which was Mrk\,463. SINFONI has a much smaller field of view, 8$''$$\times$8$''$, hence these observations only cover the central region around the dual nuclei. 

The observations were carried out using the 125$\times$250~mas spatial scale in seeing-limited mode (i.e., no adaptive optics), together with the $K$-band grating, which provides coverage from 1.95 to 2.45~$\mu$m and a spectral resolution of $\sim$4000. The observations were carried out in service mode and spread over 4 OBs: IDs 1053697 (April 17, 2014), 1053700 (May 17, 2014), 1053702 (June 18, 2014) and 1053704 (July 21, 2014). The requested weather conditions were ambient seeing better than 0.8$''$, clear skies and airmass $<$1.5. However, only OBs 1053697 and 1053700 met such conditions; the remaining ones were either taken in worse seeing and/or in the presence of thin clouds. Therefore, for this analysis we mostly focus on the first 2 OBs in order to achieve the highest possible image quality. 

The reduction of the SINFONI data was carried out using the dedicated ESO pipeline version 2.9.0 \citep{modigliani07} in {\it Reflex}. The following steps were performed: 
({\it i}) creation of a map of non-linear pixels from flat-field frames obtained with increasing exposure times; ({\it ii}) creation of a combined dark frame and hot pixel map; 
({\it iii}) creation of a master flat field from individual flat-field images; ({\it iv}) determination of the optical distortion coefficients and computation of the slitlets' distance 
table; ({\it v}) creation of a wavelength map from a set of arc lamp observations in order to carry out the wavelength calibration; ({\it vi}) stacking of individual science and 
telluric star frames. The flux calibration of the science frames was performed using the Fitting Utility for SINFONI ({\it FUS}) package developed by Dr. Krispian Lowe as
part of his PhD thesis\footnote{Available at \url{http://uhra.herts.ac.uk/bitstream/handle/2299/2449/Krispian\%20Lowe.pdf}.}, based on observations of standard stars
taken less than 2 hours away and with a difference in airmass $<$0.2 from the science data. As in the case of the VLT/MUSE data, the final combined data cubes 
were analyzed using the QFitsview tool for a first look visualization and Fluxer for the Gaussian fitting of each isolated emission lines.

\subsection{ALMA}
\label{sec:alma}

During ALMA Cycle 2, we were granted 5.1 hours of priority B time to observe a sample of four nearby dual AGN, including Mrk\,463,  in band 6 as part of program 2013.1.00525.S (PI: E. Treister). Four separate spectral windows were defined: one covering the $^{12}$CO(2-1) molecular transition, providing a 10.7 km/s spectral resolution and a 2560 km/s bandwidth and the remaining three covering the mm continuum at $\sim$220-235 GHz. The observations of Mrk\,463 were taken on August 16, 2015 between 22:27 and 23:27 UT (uid://A002/Xa830fc/X4abc) and on September 16, 2015 between 21:01 and 22:01 UT (uid://A002/Xaa305c/X129). For the August 16th observations, 34 antennae were used with baselines ranging from 43 m to 1.57 km, while on September 16th the same number of antennae were used distributed between 41 m and 4.5 km baselines. The total on-target time in each observation was 25.7 minutes. Data reduction and analysis were carried out using the Common Astronomy Software Applications (CASA) version 4.5.0.

The requested 3-$\sigma$ sensitivity of the observations was 0.58 mJy/beam per $\sim$10.7 km/s channel. This requirement was not quite met, as the early
processing carried out by the North America ALMA Science Center yielded a sensitivity of 0.78 mJy/beam on the $^{12}$CO(2-1) spectral window, 34\% higher than 
the goal value. This value was obtained using a Brigg Robust image weighting \citep{briggs95} with a robust parameter of 0.5, that provides a balance between a 
smaller beam but with a lower sensitivity. Instead, we decided to re-process the ALMA data using a natural weighting scheme, which delivers a higher sensitivity at 
the expense of a slightly worse spatial resolution. Using this method, we achieved a sensitivity of $\sim$0.6 mJy/beam at 3-$\sigma$ in the $^{12}$CO(2-1) line, thus 
much closer to our goal flux limit. The resulting beam size is 0.3$''$$\times$0.17$''$.

\subsection{Astrometric Calibration}
\label{sec:astrometry}

In order to be able to compare emission at different wavelengths from the instruments described above it is critical to ensure that all the datasets share
the same astrometric reference point. This is potentially difficult given the relatively small overlap in structures at different wavelengths. The ALMA absolute
astrometric calibration is better than $\sim$0.1$''$ as reported by \citet{gonzalez-lopez17}. 

Using the {\it Chandra} X-ray data for this system, \citet{bianchi08} located 
the position of the two AGN to within 0.5$''$. In particular, we derived the exact position using the hard X-ray band, 2--8 keV, 
as in this case most of the emission can be attributed to the AGN.

\section{Morphologies}
\label{sec:morph}

Mrk\,463 was morphologically classified as a peculiar/interacting galaxy based on early optical imaging \citep{adams77,petrosian78}. Deeper CCD imaging by \citet{hutchings89}, which included a narrow-band filter covering the redshifted [O{\sc iii}] 5007\AA~emission lines, revealed extended emission in the Mrk\,463E nucleus and a bright knot $\sim$10$''$ south of the nucleus that are not visible in the optical continuum. Radio continuum observations at 6 and 20 cm performed by \citet{mazzarella91} show that the radio emission is aligned with the bi-polar [O{\sc iii}] conical flows and the southern knot, strongly suggesting that they are connected to the Mrk\,463E nuclear region.

\begin{figure*}
\vspace{-0.0in}
\begin{center}
\hglue-0.2cm{\includegraphics[width=7in]{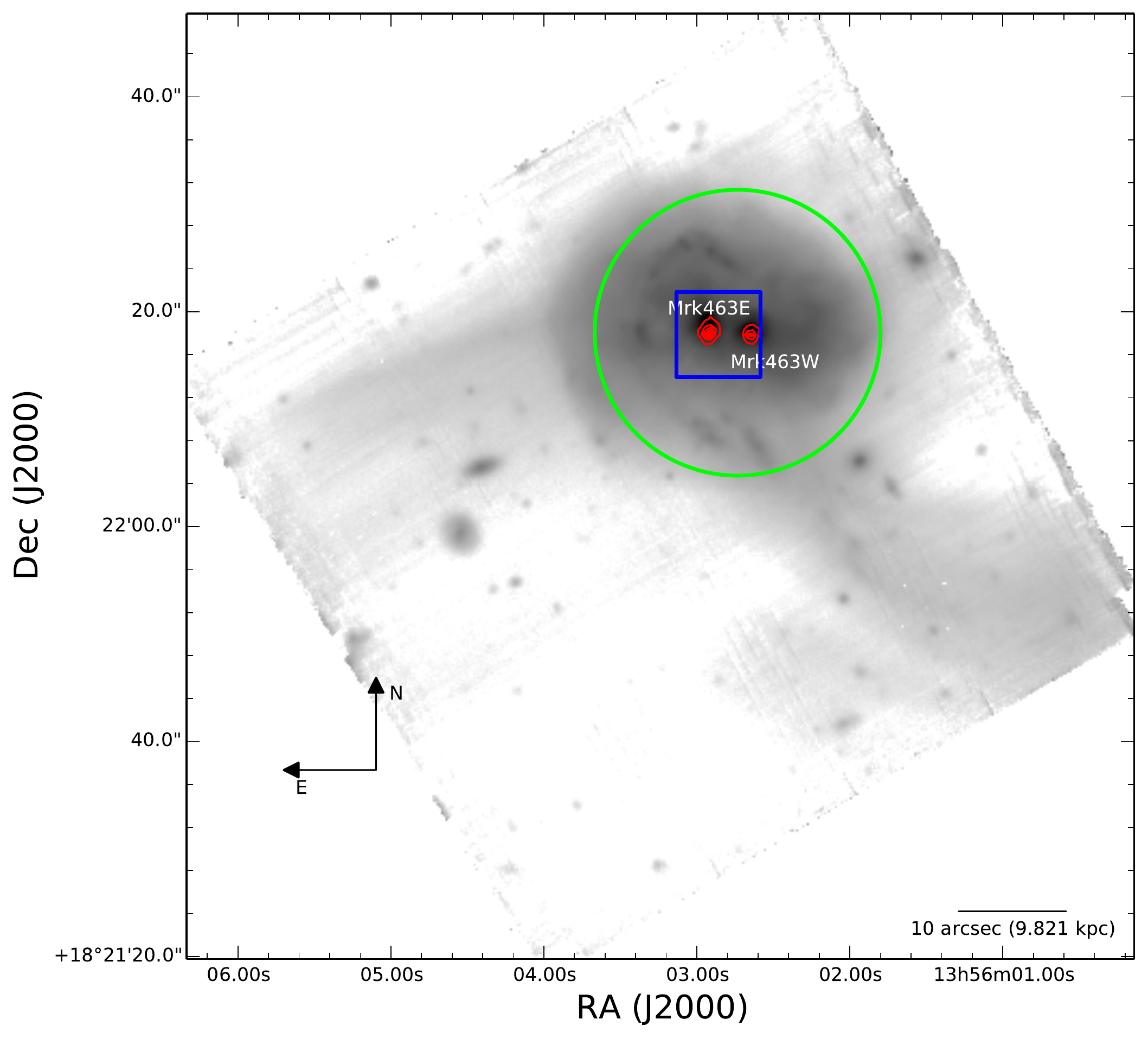}}
\end{center}
\vspace{-0.2in}
\caption{Reconstructed white filter VLT/MUSE image for Mrk\,463 covering the entire wavelength range from $\sim$4800 to $\sim$9300 \AA. North is up and east to 
the left. Pixel scale is the natural 0\farcs2/pixel. Spatial resolution is seeing limited and $\sim$0\farcs7. {\it Red contours} show the {\it Chandra} hard X-ray, 2-8 keV, 
emission while the {\it green} circle and {\it blue} rectangle highlight the regions covered by the ALMA and VLT/SINFONI observations, respectively. \label{MUSE_cont}}
\end{figure*}

Taking advantage of the MUSE IFU data, in Fig.~\ref{MUSE_cont} we present a reconstructed optical continuum image of the Mrk\,463 system, ranging from 
$\sim$4800 to $\sim$9300\AA. This image covers the entire MUSE $\sim$1$\times$1 arcmin$^2$ field of view, roughly an order of magnitude larger than previous IFU 
observations of this source \citep[e.g.,][]{chatzichristou95}. Both nuclei, Mrk\,463E and Mrk\,463W, are clearly visible on the image, together with the southern knot, which 
is almost exclusively seen in [O{\sc iii}] emission, as discussed in \S\ref{sec:morph_optlines}. Additionally, we can see relatively bright emission clumps, most 
prominently to the north-east of the Mrk\,463E galaxy, which we will later, in the following subsection, associate with off-nuclear star-forming regions.

\subsection{Optical Atomic Transitions}
\label{sec:morph_optlines}

In order to obtain line flux maps from the MUSE data cubes we perform Gaussian fits to the He II 4685\AA, $H\beta$ 4861\AA, [O{\sc iii}] 5007\AA, 
$H\alpha$ 6563\AA, [N{\sc ii}] 6548,6583\AA~ and [S{\sc ii}] 6717,6731\AA~ features. The line fitting procedure is as follows: First, the continuum 
emission in each pixel is subtracted by performing a fit using a second order polynomial function between 5000\AA~and 8000\AA. Using this broad wavelength range and 
a relatively small polynomial order allows to obtain a good fit for the spectral continuum in the region of interest while remaining unaffected by the presence of even 
prominent emission or absorption features. Then, we performed a Voronoi tesselation of the cube by demanding a minimal signal-to-noise ratio of 10 in the 
$H\alpha$/[N{\sc ii}] complex in each bin. While this choice has no impact on the central regions of the system, where this condition is met basically on every 
pixel, it is important in the outskirts. We then assume Gaussian profiles for all emission lines. Given that it is a strong and well-isolated feature, we use 
the [O{\sc iii}] 5007\AA~line as a template in order to define the widths and relative wavelength offsets for all the subsequent narrow components. We further 
allow for a secondary line component at a different velocity, also using the [O{\sc iii}] 5007\AA~line as template and adding extra Gaussians with both the line centers 
and widths as free parameters to account for the possible presence of broad $H\alpha$ and $H\beta$ emission. In each case, a component is only considered if the 
peak is detected at a signal-to-noise greater than 3. 

\begin{figure*}
\vspace{-0.0in}
\begin{center}
\hglue-0.0cm{\includegraphics[width=3.4in]{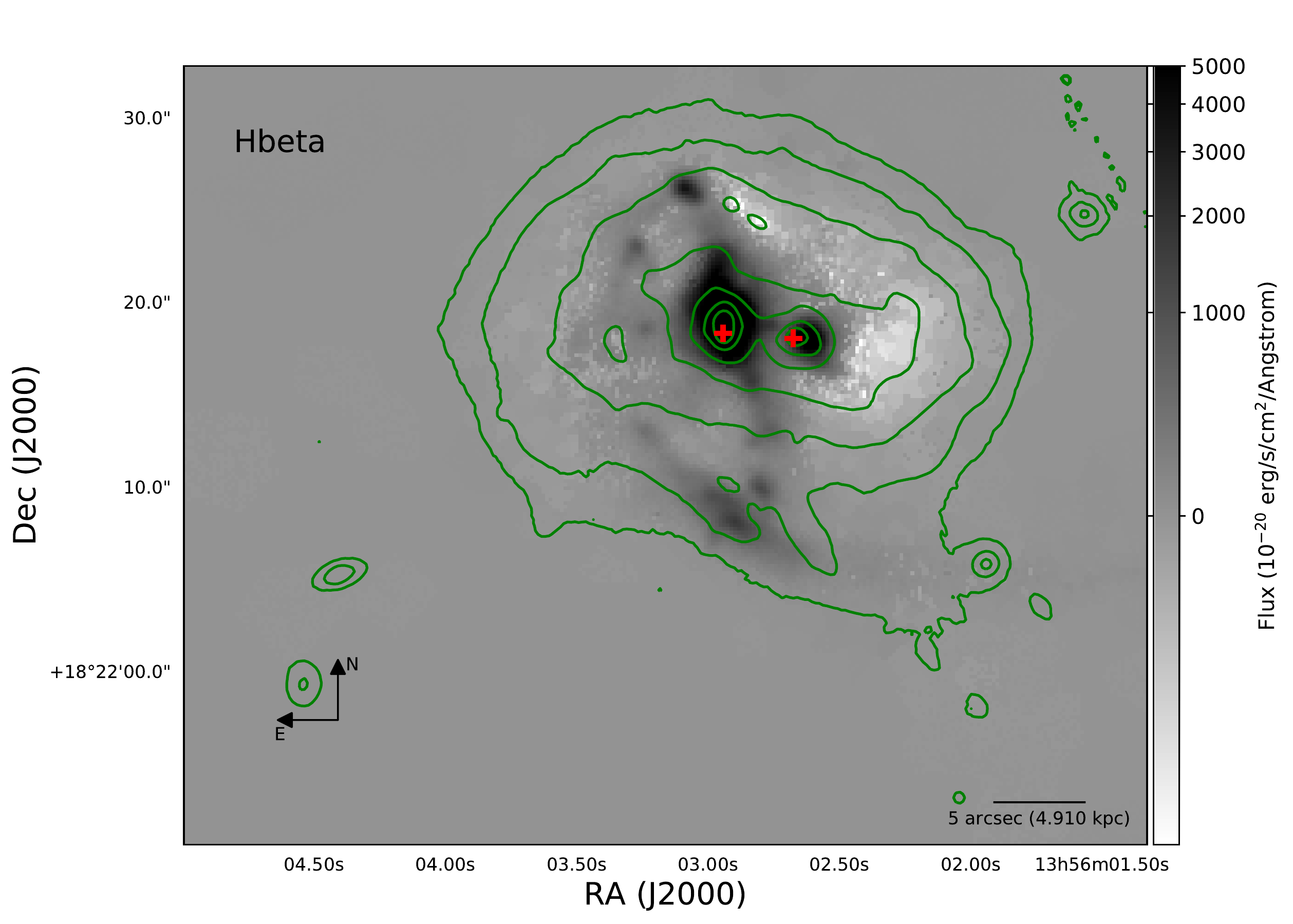}}\hfill
\hglue-0.0cm{\includegraphics[width=3.4in]{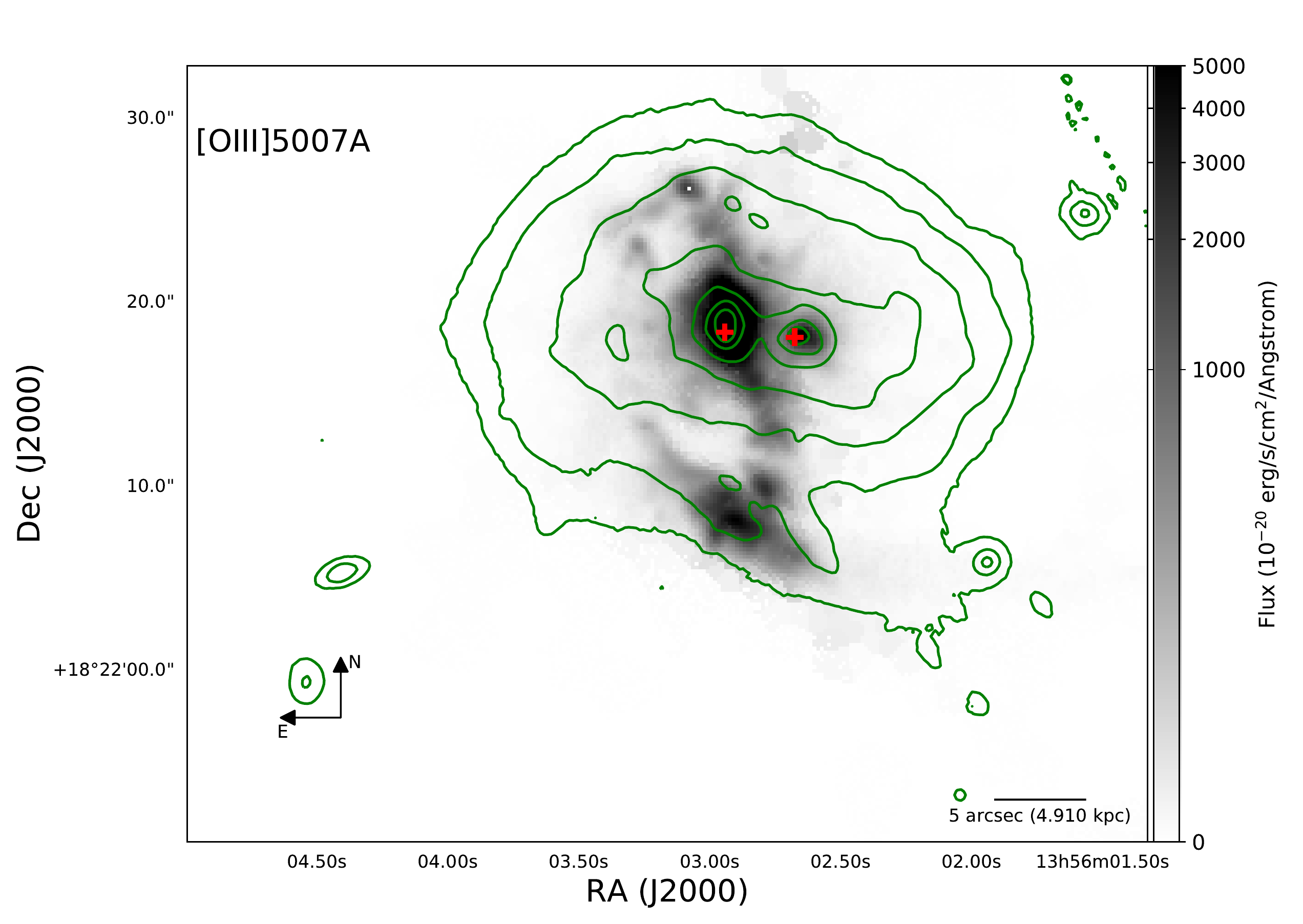}}
\end{center}
\vspace{-0.1in}
\caption{Flux distributions (grey scale) for the Mrk\,463 system in the $H\beta$ 4861\AA~({\it left panel}, ranging from -5$\times$10$^{-18}$ to 
5$\times$10$^{-17}$ erg/cm$^2$/s to include the absorption features) and [O{\sc iii}] 5007\AA~({\it right panel}) atomic transitions. Fluxes, given in both cases 
in units of 10$^{-20}$erg/cm$^2$/s, were measured in each pixel by summing the contribution from the Gaussian fit for each narrow component, with the 
profile given by the [O{\sc iii}] 5007\AA~line, after subtracting the surrounding continuum. For reference, the {\it red crosses} show the location of the 
X-ray sources, while the {\it green contours} correspond to the optical continuum flux.\label{img_Hbeta_OIII}}
\end{figure*}

\begin{figure*}
\vspace{-0.1in}
\begin{center}
\hglue-0.0cm{\includegraphics[width=3.4in]{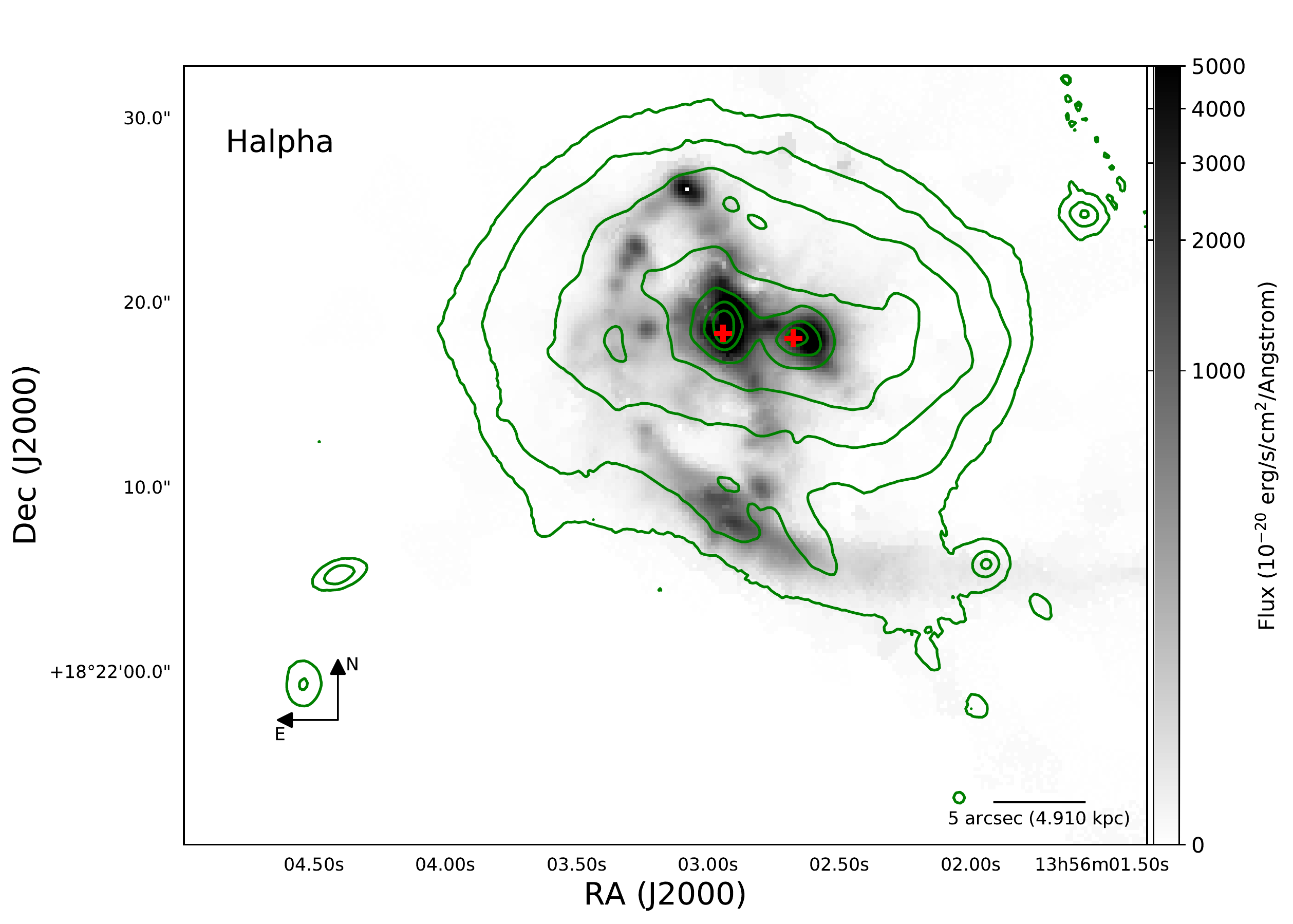}}\hfill
\hglue-0.0cm{\includegraphics[width=3.4in]{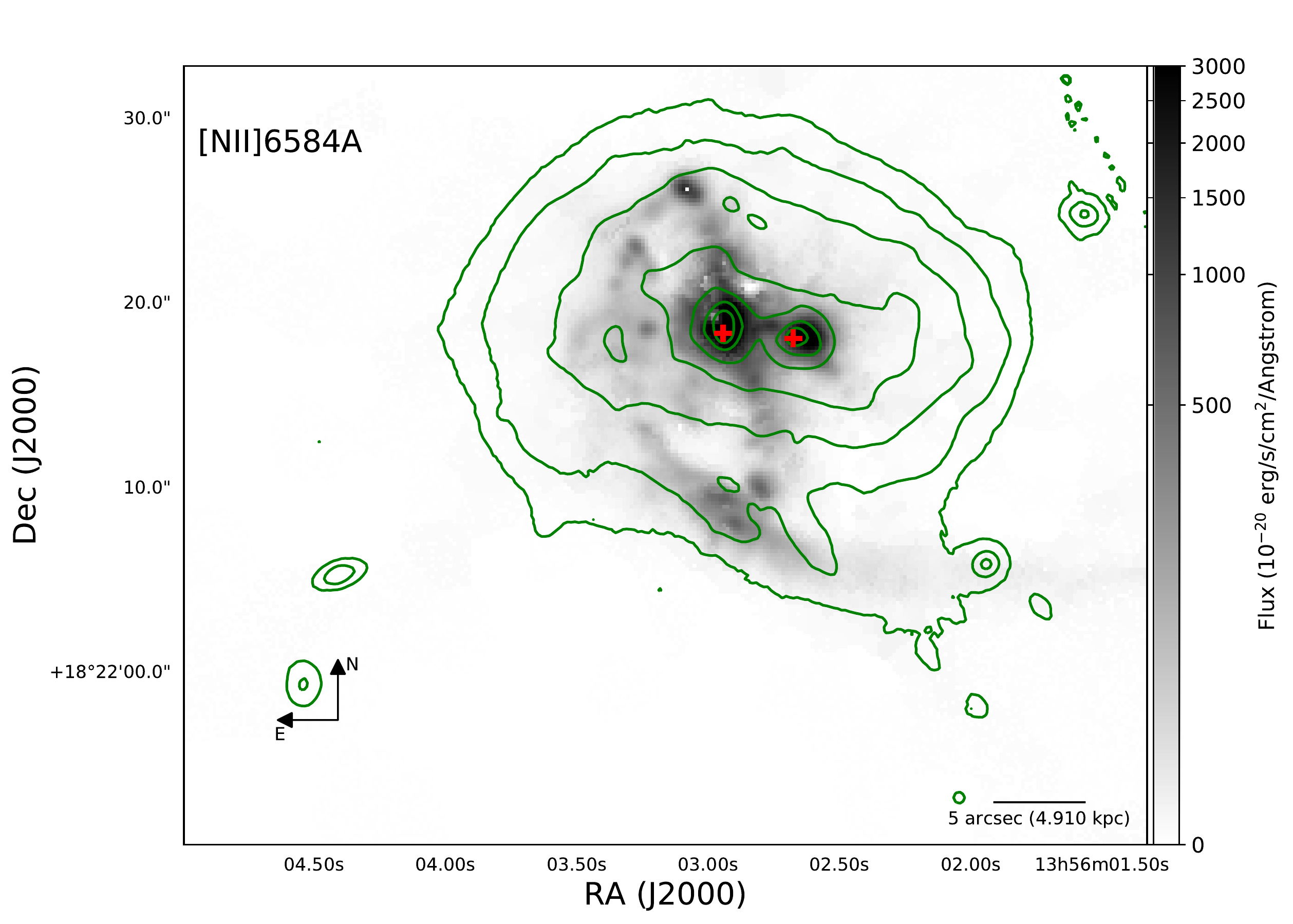}}
\end{center}
\vspace{-0.1in}
\caption{Emission maps (grey scale) for the $H\alpha$ 6563\AA~({\it left panel}) and the [NII] 6583\AA~({\it right panel}) transitions obtained following the 
procedure described in the text. Contours, symbols and units are the same as in Fig.~\ref{img_Hbeta_OIII}. \label{img_Halpha_NII}}
\end{figure*}

The resulting images are presented in Figs.~\ref{img_Hbeta_OIII} and \ref{img_Halpha_NII}. In all of these cases, the total flux for all the line components is
shown. Some of the same key features seen in Fig.~\ref{MUSE_cont} and discussed in the previous section are clearly visible in these images as well. However, 
other remarkable components are now detected as well. Particularly important are the detections of extended $H\alpha$ gas to the northwest of the system and 
the ``stream'' of $H\alpha$ blobs to the southwest, seen in Fig.~\ref{img_Halpha_NII}.  Additionally, in the $H\beta$ 4861\AA~map we can identify an absorption 
region to the west of the Mrk 463W nucleus. 

\begin{figure*}
\vspace{-0.0in}
\begin{center}
\hglue-0.0cm{\includegraphics[width=3.4in]{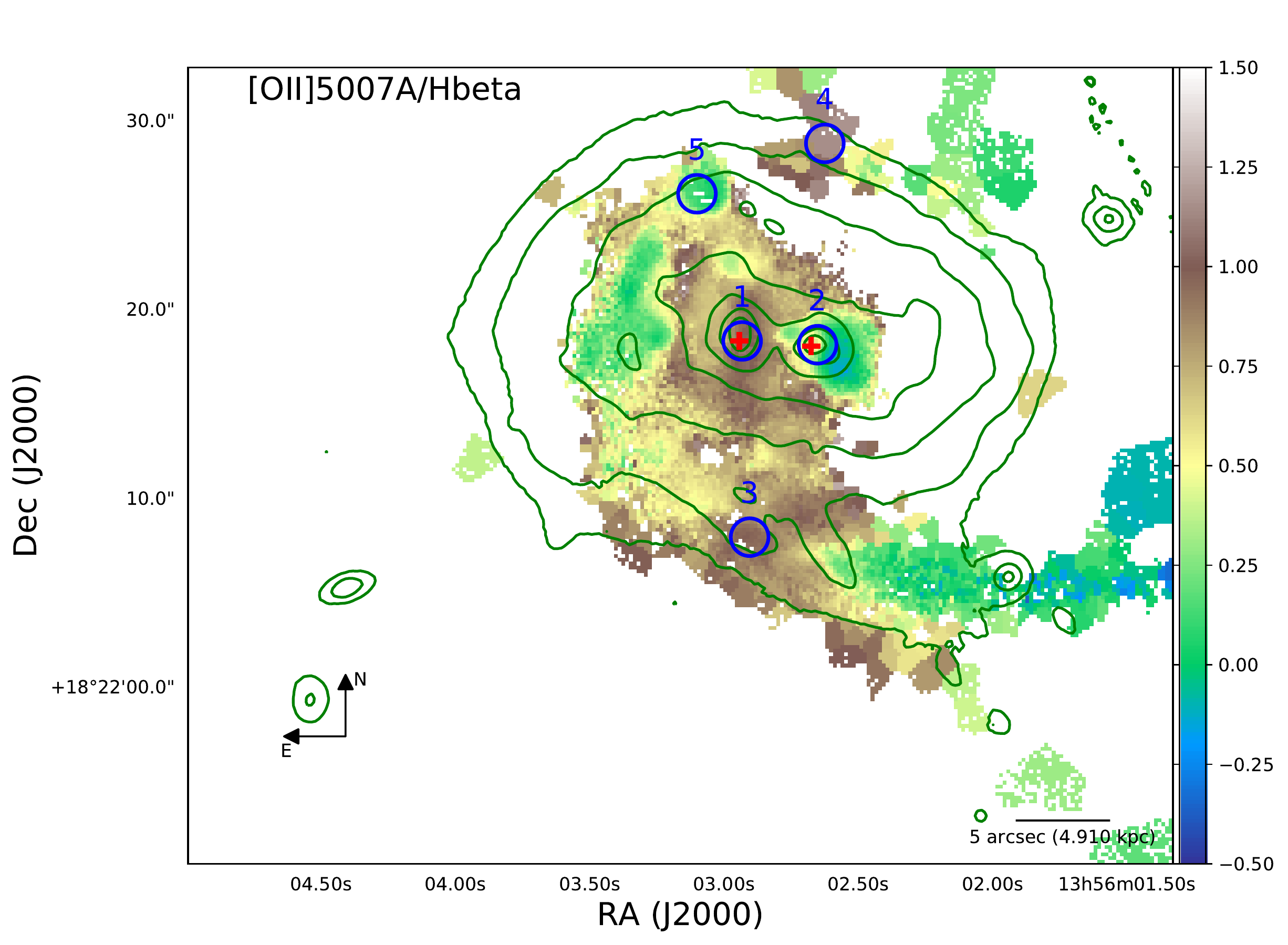}}\hfill
\hglue-0.0cm{\includegraphics[width=3.4in]{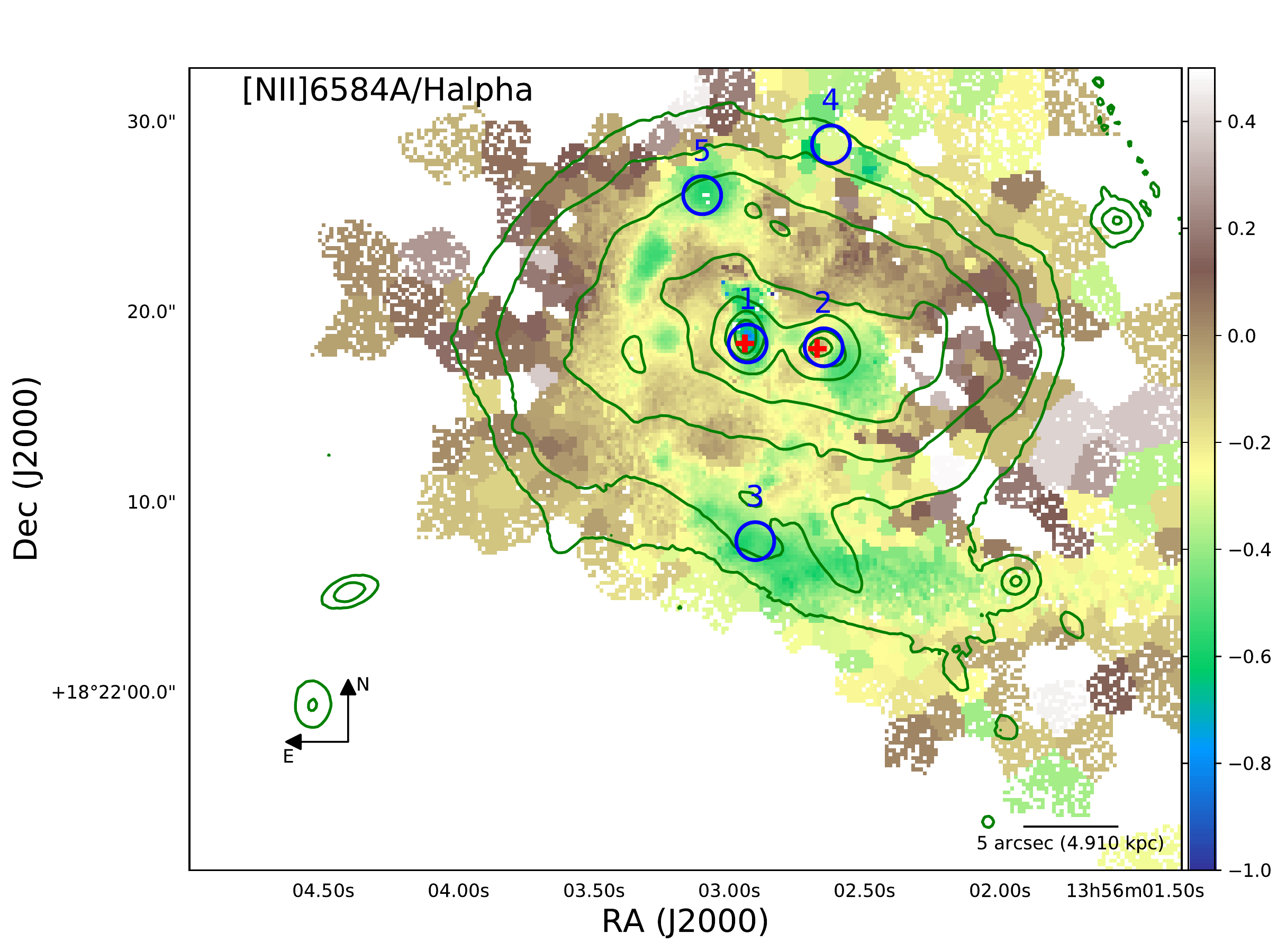}}
\end{center}
\vspace{-0.1in}
\caption{Map of the flux ratio between [O{\sc iii}] 5007\AA~and $H\beta$ 4861\AA ({\it left panel}) and [NII] 6584\AA~ to $H\alpha$ 6563\AA~({\it right panel})
emission lines for the Mrk\,463 system. Only pixels with a signal-to-noise ratio $>$3 in each line independently have been included. {\it Green contours} show 
the position of the integrated optical emission for reference, while the {\it red crosses} mark the hard X-ray emission. The five numbered {\it blue circles} show the 
regions in which the emission line ratios were measured and presented in Fig.~\ref{bpt}.\label{img_line_ratios}}
\end{figure*}

\begin{figure}
\begin{center}
\hglue-0.0cm{\includegraphics[width=3.5in]{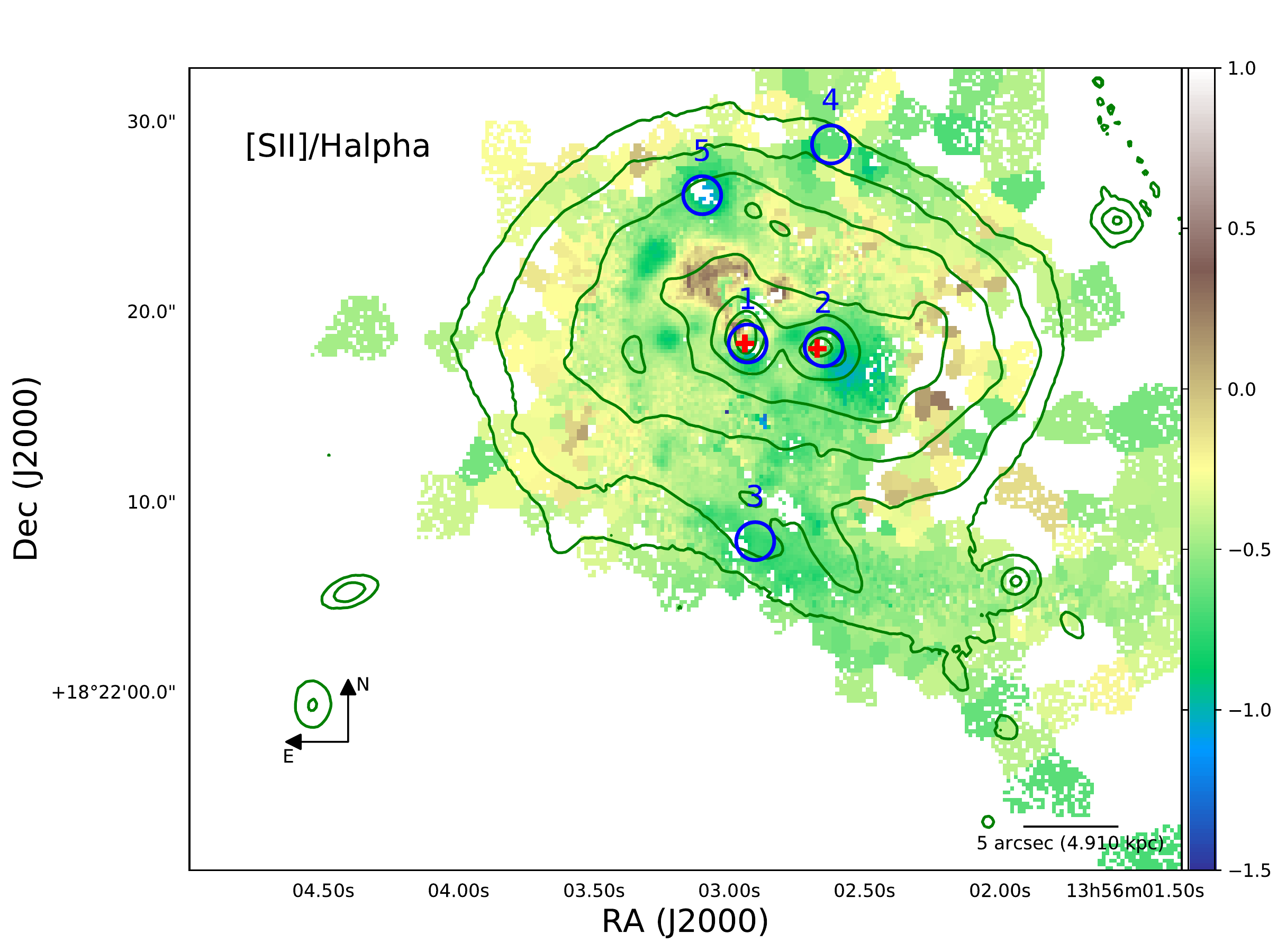}}
\end{center}
\vspace{-0.2in}
\caption{[SII] to $H\alpha$ flux ratio for the Mrk 463 system. Contours and symbols are the same as in Fig.~\ref{img_line_ratios}.}
\label{img_SII_Halpha}
\end{figure}

\begin{figure*}
\vspace{-0.1in}
\begin{center}
\hglue-0.0cm{\includegraphics[width=3.4in]{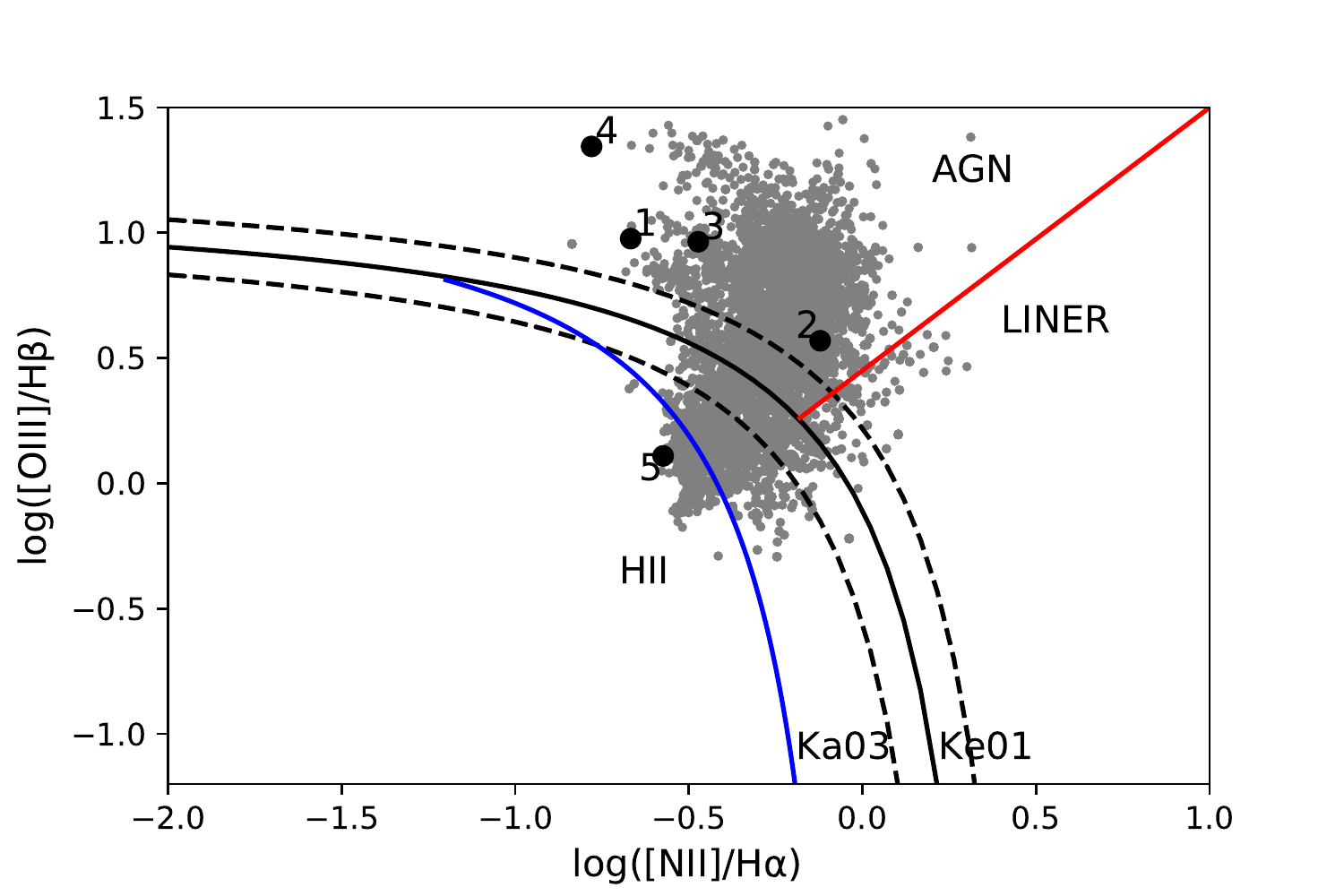}}\hfill
\hglue-0.0cm{\includegraphics[width=3.4in]{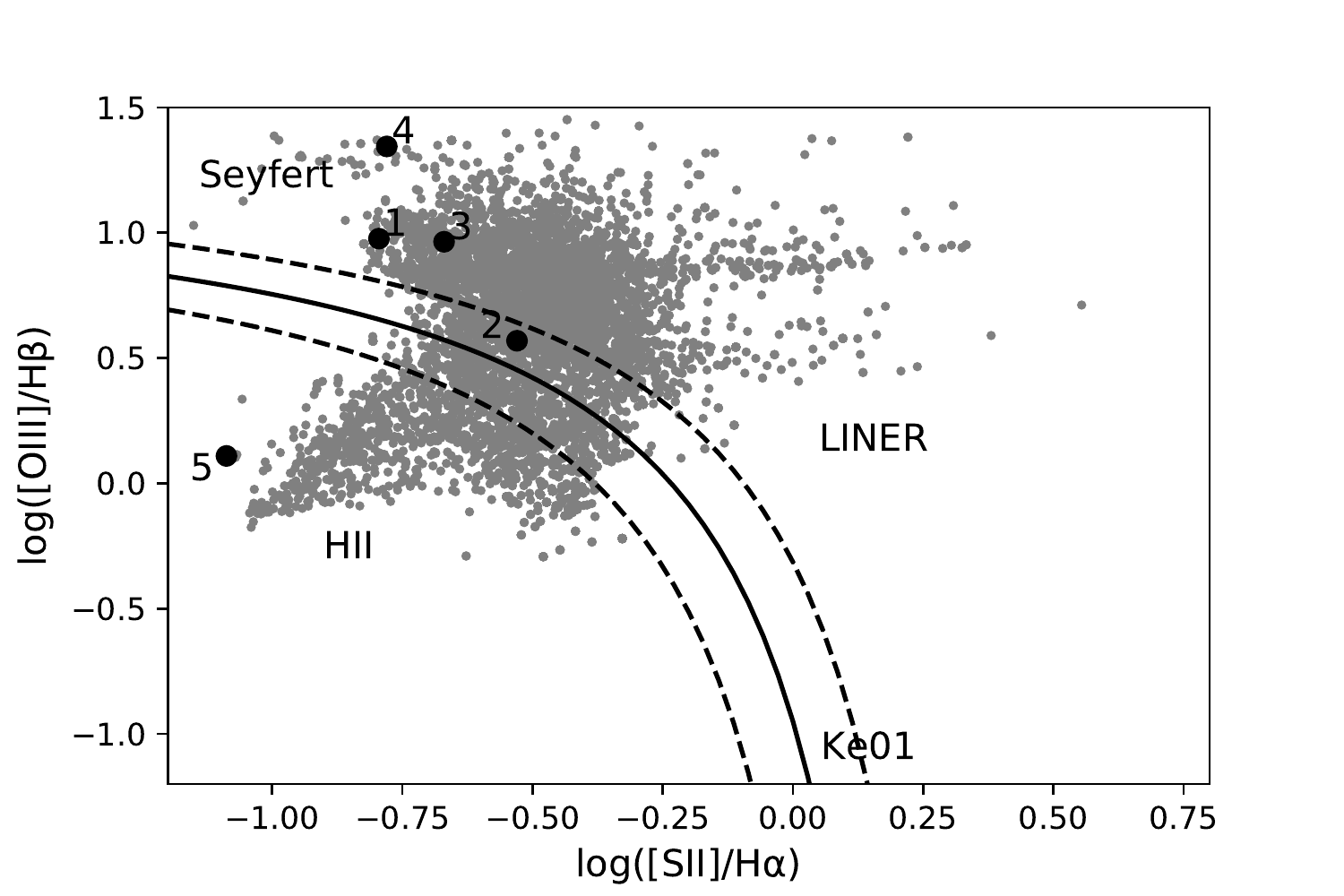}}
\end{center}
\vspace{-0.2in}
\caption{[NII]/$H\alpha$ versus [O{\sc iii}]/$H\beta$ (left panel) and [SII]/$H\alpha$ versus [O{\sc iii}]/$H\beta$ (right panel) diagrams for the five regions 
highlighted in Figure~\ref{img_line_ratios}. The {\it black solid curves} show the separation between star-forming (HII) and active galaxies proposed by \citet{kewley01}, 
while the adjacent {\it dashed curves} represent $\pm$0.1 dex of this relation, corresponding to the estimated error for this threshold. The {\it red solid line} 
on the left panel shows the separation between AGN and LINERs proposed by \citet{schawinski07}, while the {\it blue solid curve} marks the classification
threshold presented by \citet{kauffmann03}. The background {\it grey circles} in both panels show the locations in this diagram for all the Voronoi bins in 
Mrk463 with detected emission lines. The {\it black circles} show the position of the highlighted regions in the system. As can be seen, regions 1-4 are clearly in 
the AGN zone, while the position 5 can be classified as a star forming region.\label{bpt}}
\end{figure*}

The flux ratios of these emission lines can be used in order to determine the nature of the ionization source \citep[e.g.,][]{baldwin81,veilleux87,kewley06}. 
Figures~\ref{img_line_ratios} and \ref{img_SII_Halpha} show the maps for the [O{\sc iii}]/$H\beta$, [SII]/$H\alpha$, and [NII]/$H\alpha$ ratios, 
computed using the sum of the flux in the narrow components for each emission line. According to \citet{kewley06}, any source/region with 
[O{\sc iii}]/$H\beta$$>$10 is nearly guaranteed to be dominated by AGN ionization, independent of the value of [NII]/$H\alpha$ or [SII]/$H\alpha$. While 
shocks appear to be widespread in major galaxy mergers, they typically have lower [O{\sc iii}]/$H\beta$ ratios, $\sim$1-5, compared to active 
nuclei \citep{rich11,rich15}. Hence, just from examination of Fig.~\ref{img_line_ratios} we can see several potential AGN-dominated regions. In order to 
investigate the nature of the ionization energy source(s), we selected five outstanding zones in the system, highlighted in Figures~\ref{img_line_ratios} 
and \ref{img_SII_Halpha}: the two nuclear regions,  the southern emission line region, the northern clump, and a representative $H\alpha$ emitting region 
in the northwest. The location of these five regions on the [O{\sc iii}]/$H\beta$ versus [NII]/$H\alpha$ and [SII]/$H\alpha$ diagnostic diagrams is shown in 
Figure~\ref{bpt}. As can be clearly seen, both nuclei, the southern emission line region and the northern clump are consistent with being dominated by the AGN 
energy output. In contrast, region \#5 appears to fall on the star-forming locus. These diagrams indicate that the 
influence of the AGN on the surrounding material is widespread and can be detected at large distances, even $\sim$10 kpc away from the nuclei.

\begin{figure*}
\vspace{-0.1in}
\begin{center}
\hglue-0.0cm{\includegraphics[width=3.4in]{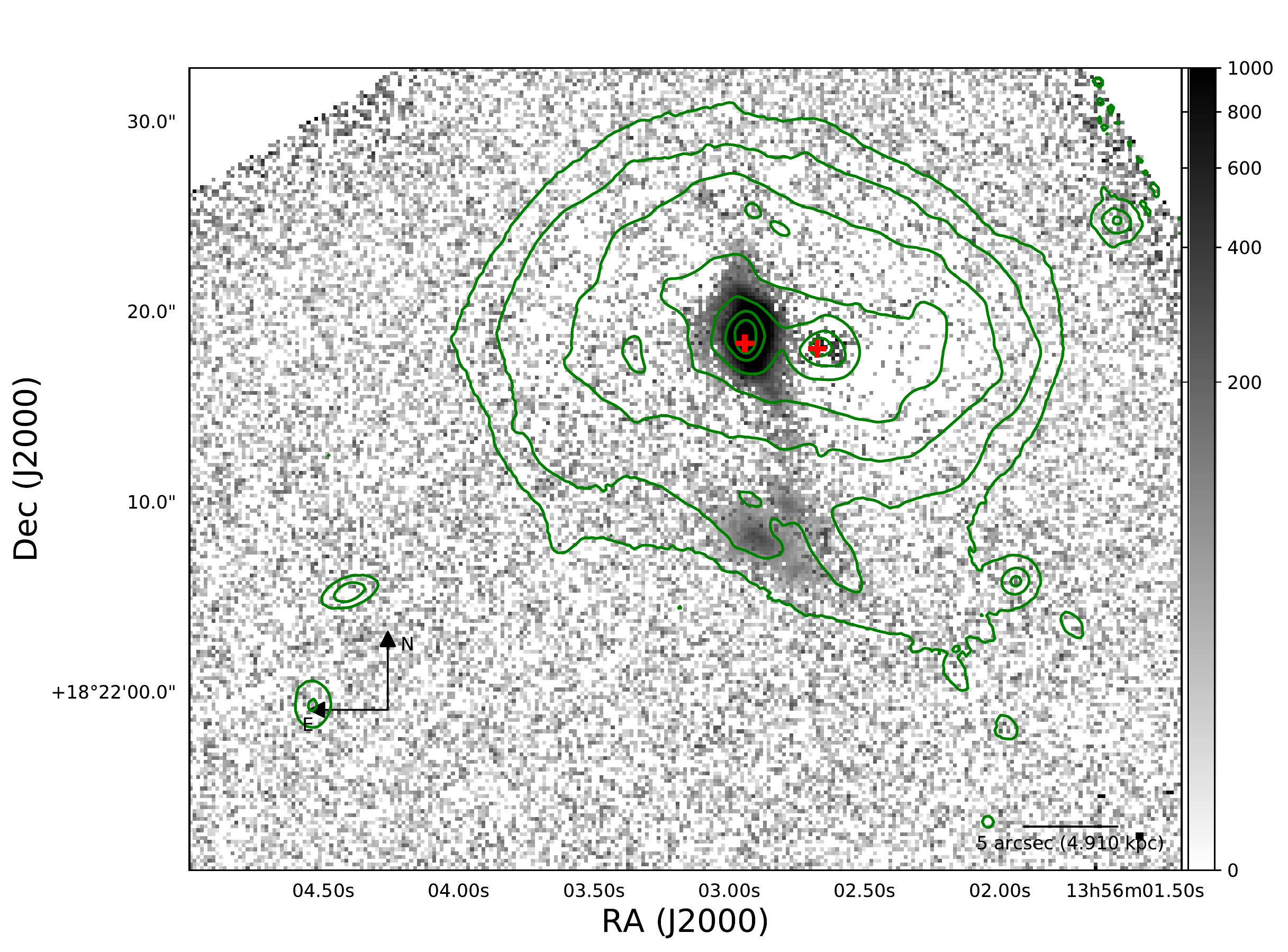}}\hfill
\hglue-0.0cm{\includegraphics[width=3.4in]{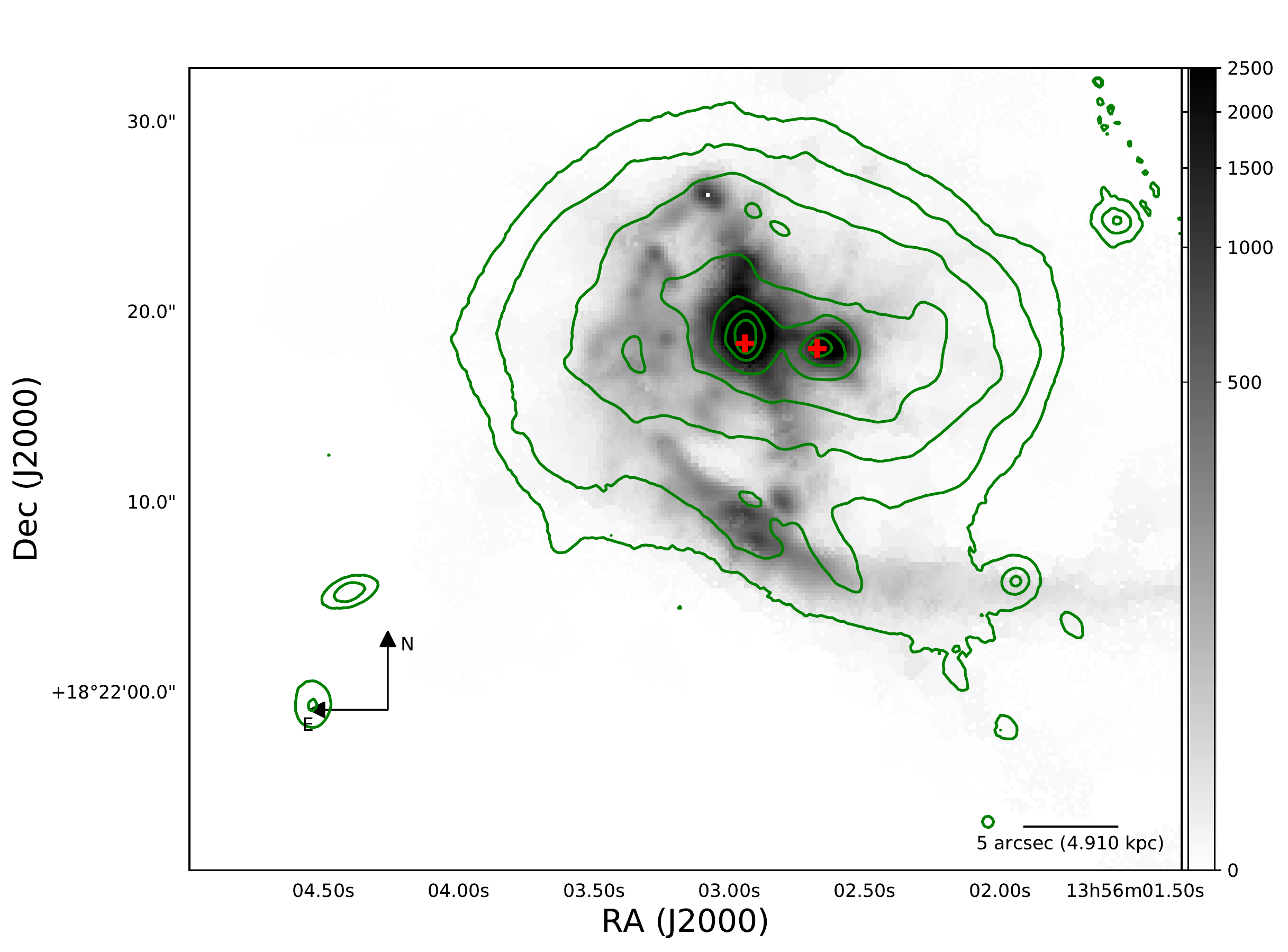}}
\end{center}
\vspace{-0.1in}
\caption{Flux maps for the He II 4685\AA~({\it left panel}) emission line and for the [SII] 6717/31\AA~ doublet ({\it right panel}) obtained from the 
VLT/MUSE IFU data cubes.  Contours, symbols and units are the same as in Fig.~\ref{img_Hbeta_OIII}. Both maps were obtained by performing Gaussian fits to the 
emission lines. \label{img_HeII_SII}}
\end{figure*}

Further hints about the  nature of the ionization sources in this system can be obtained by exploring emission lines with higher ionization energies. To this end, 
Fig.~\ref{img_HeII_SII} shows the emission maps for the He II 4685\AA~and the [SII] 6717/31\AA~lines, which have ionization energies of 54.4 eV and 23.3 eV respectively. 
As can be seen in the figure, highly ionized lines, in particular He II, are much more concentrated, mostly on the Mrk\,463E nucleus and the emission line regions, 
while Mrk\,463W is barely detected. As presented by \citet{shirazi12} and later by \citet{bar16}, the He II 4685\AA~emission and the He II/H$\beta$ ratio, can be 
used to identify AGN even in sources missed by classification methods based on emission lines at lower ionization energies \citep{sartori15}. According to \citet{shirazi12}, 
sources or regions with $\log$(He II/$H\beta$)$>$-1 can be classified as AGN dominated. This is the case for the Mrk\,463E nucleus and its surrounding ionization cone, 
for the southwest emission line region but not for the Mrk\,463W nucleus. This suggests that the AGN in Mrk\,463E is energetically more important for the system than 
the one in Mrk\,463W.

\subsection{Near-IR maps}
\label{sec:morph_near-IR}

\begin{figure*}
\vspace{-0.1in}
\begin{center}
\hglue-0.0cm{\includegraphics[width=3.4in]{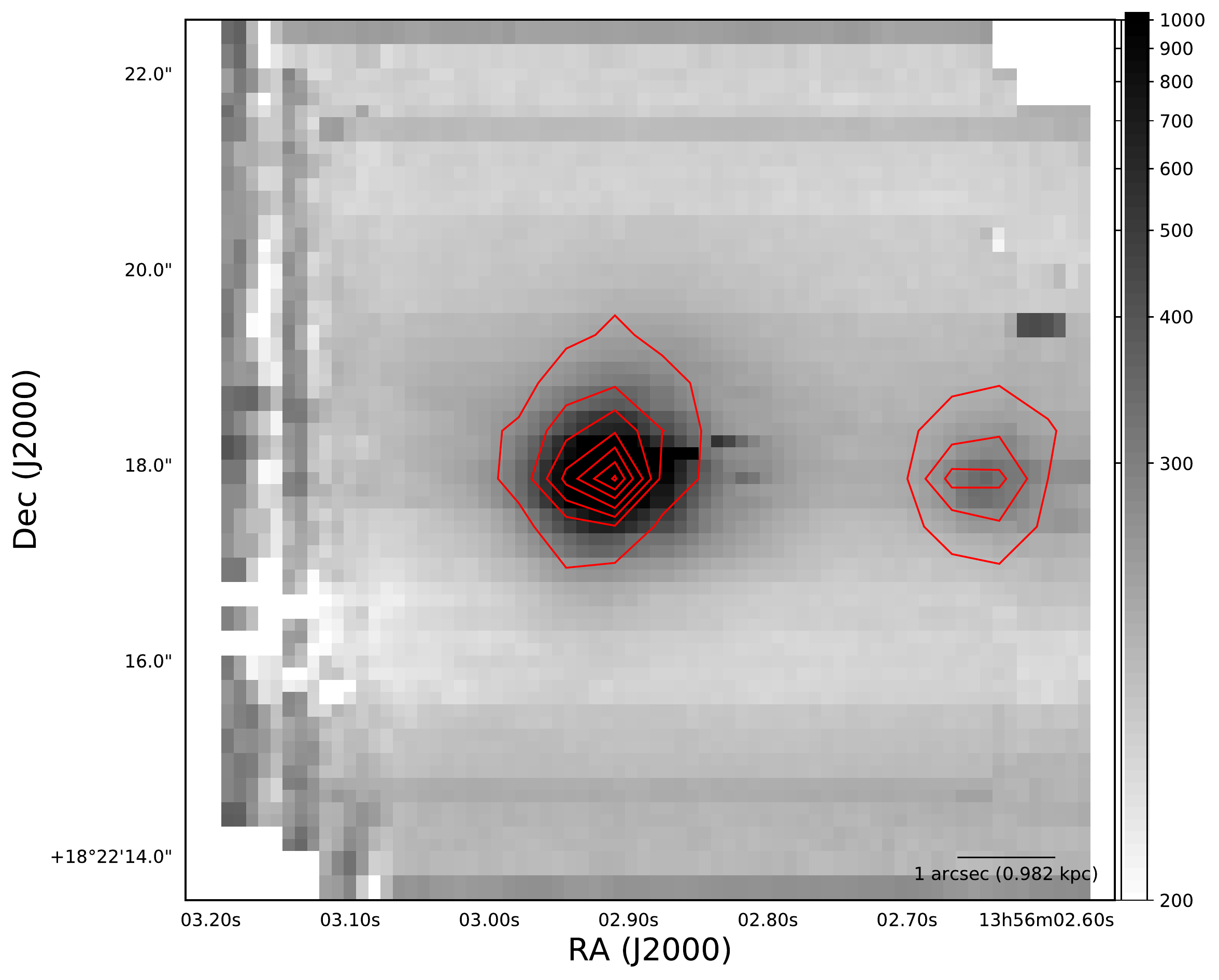}}\hfill
\hglue-0.0cm{\includegraphics[width=3.4in]{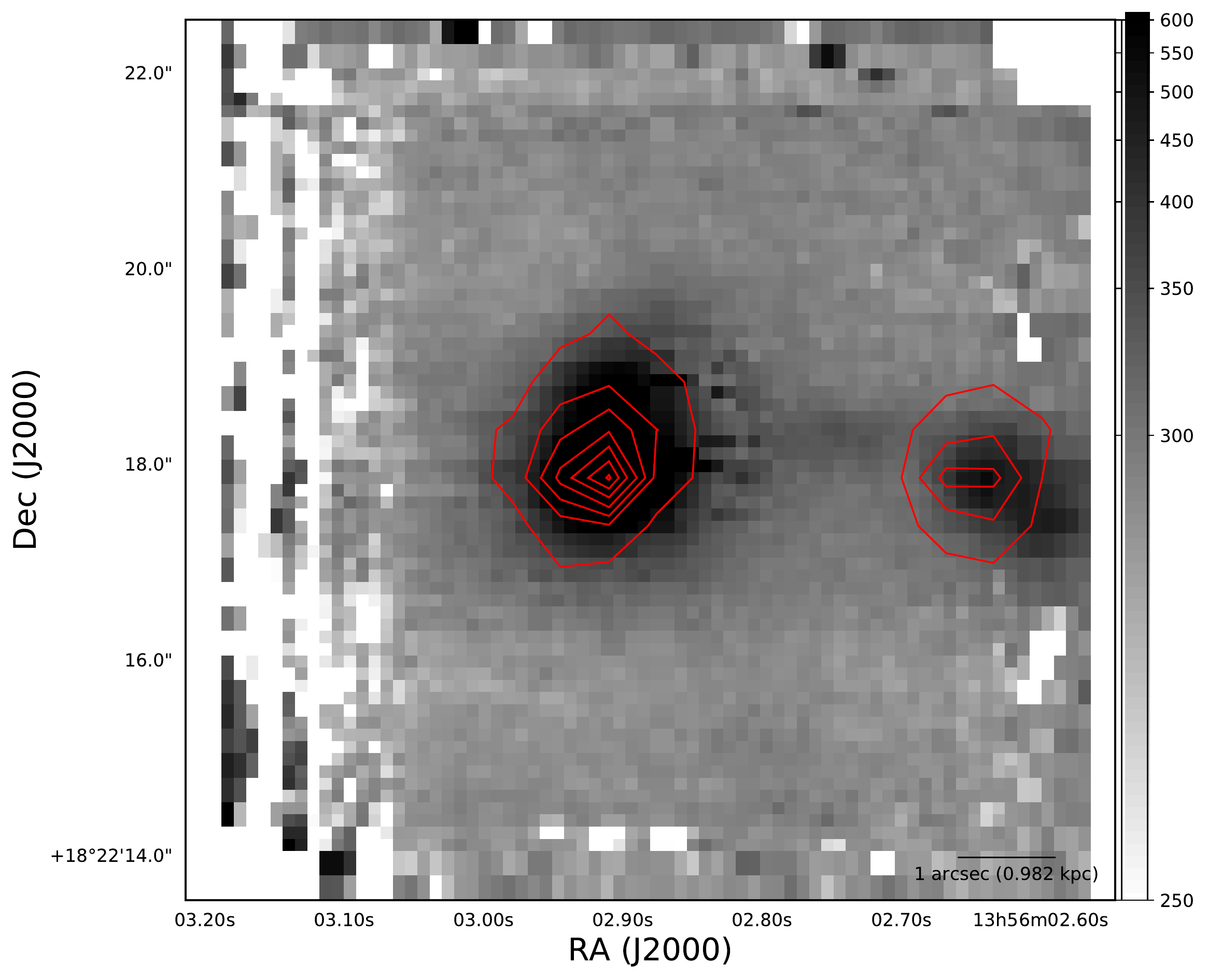}}
\end{center}
\vspace{-0.1in}
\caption{{\it Left panel:} Mrk\,463 near-IR continuum at $\sim$2.2$\mu$m. {\it Right panel}: emission map
for the Pa$\alpha$ hydrogen line at 1.8751 $\mu$m. The VLT/SINFONI cubes were used to create these maps. The near-IR continuum was obtained by 
collapsing the data cube between 2.1 to 2.4~$\mu$m in the observed frame. The Pa$\alpha$ map was created by fitting a Gaussian function to the 
emission line after subtracting the surrounding continuum. Red contours in both cases show the hard X-ray emission detected by Chandra.
\label{img_kband_paalpha}}
\end{figure*}

Although the VLT/SINFONI near-IR IFU data only cover a smaller $\sim$8$''$$\times$8$''$ region at the center of the Mrk\,463 system, as shown
in Fig.~\ref{MUSE_cont}, they are an important complement to the MUSE optical maps. Figure~\ref{img_kband_paalpha} shows the near-IR continuum between 2.1 and 2.4 
$\mu$m and the continuum-subtracted integrated Pa$\alpha$ hydrogen emission line at 1.8751 $\mu$m in the rest frame. For this and all near-IR emission lines, the 
continuum subtraction was carried out by performing a simple first order polynomial fit to the adjacent spectral regions on each side of the line. The line fitting was carried
out assuming a single Gaussian functional form, which was visually deemed appropriate and reasonable considering that the SINFONI data only cover the central
region of the system. No cut in signal to noise ratio of the resulting line was used in constructing the map. Both nuclei are clearly detected on the continuum map, 
showing a very similar morphology to the optical continuum image presented in Fig.~\ref{MUSE_cont}. The Pa$\alpha$ map, in contrast, reveals 
new components. In particular, we can marginally detect an emission blob between the two nuclei and another emitting region to the west of the Mrk\,463W nucleus. While 
both of these regions are also visible in the H$\alpha$ map, they are not detected in [O{\sc iii}], suggesting that they are star-forming regions. However,
as it will be later shown in section~\ref{sec:opt_ext}, the region between the nuclei is subject to moderate extinction, $A_{H\alpha}$$\sim$2, and therefore it is possible
that the lack of detection of [O{\sc iii}] in this area is due to obscuration.

The Pa$\alpha$ emission line provides a measurement of the star formation rate. Using the SINFONI cube together with the conversion from Pa$\alpha$ luminosity to
star formation rate provided by \citet{rieke08}, we derive values of 26.5 M$_\odot$/yr for the East nucleus and 0.75 M$_\odot$/yr for the West one. These star 
formation rates derived from the Pa$\alpha$ line are fully consistent with those of 30 M$_\odot$/yr for Mrk 463E and $<$10 M$_\odot$/yr for Mrk 463W reported 
by \citet{evans02}. However, it is important to point out that as shown in Fig.~\ref{bpt}, both nuclei appear to be dominated by the AGN emission, at least for the
optical lines, and thus these SFR estimates can be considered as upper limits.

\begin{figure*}
\vspace{-0.1in}
\begin{center}
\hglue-0.0cm{\includegraphics[width=3.4in]{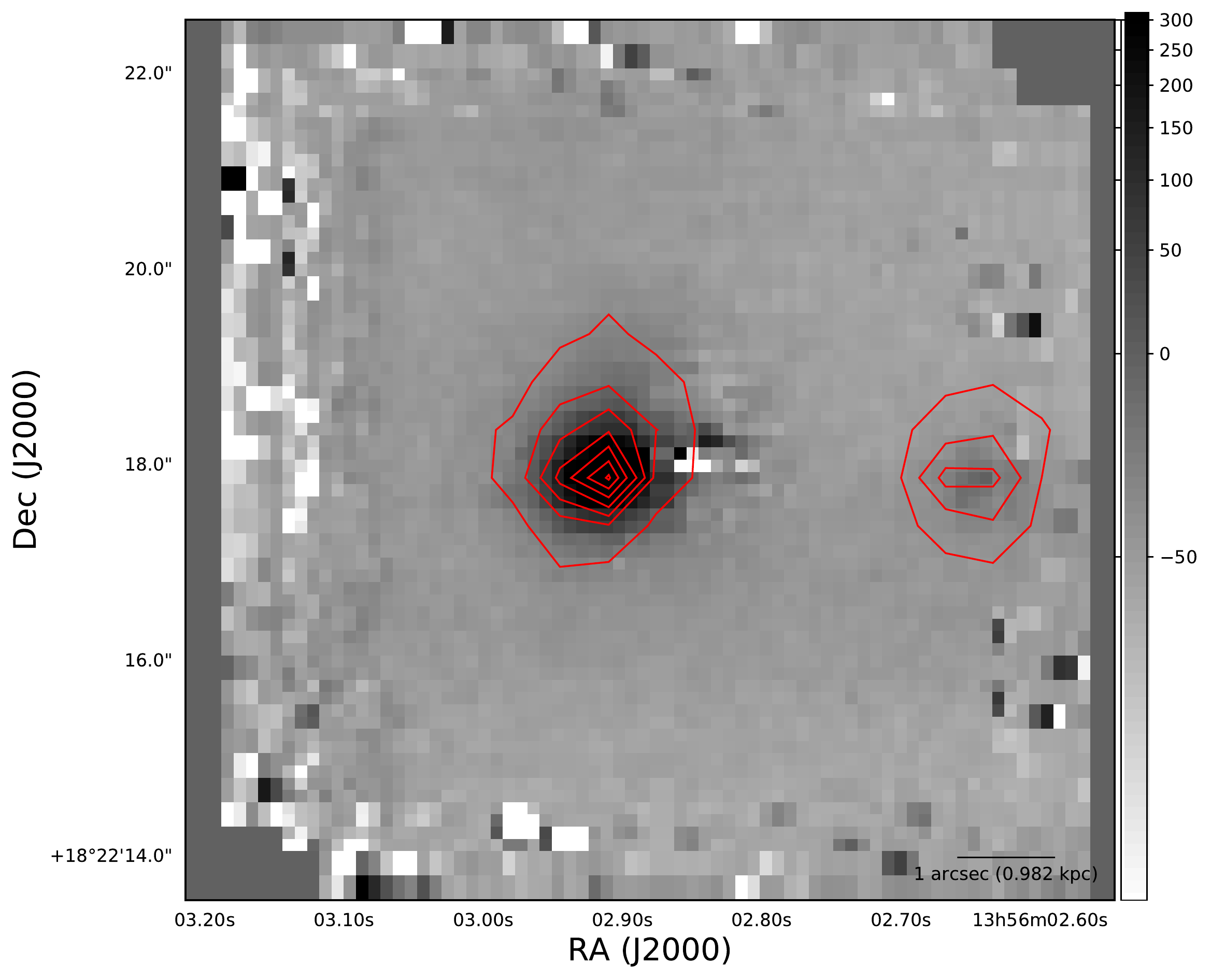}}\hfill
\hglue-0.0cm{\includegraphics[width=3.4in]{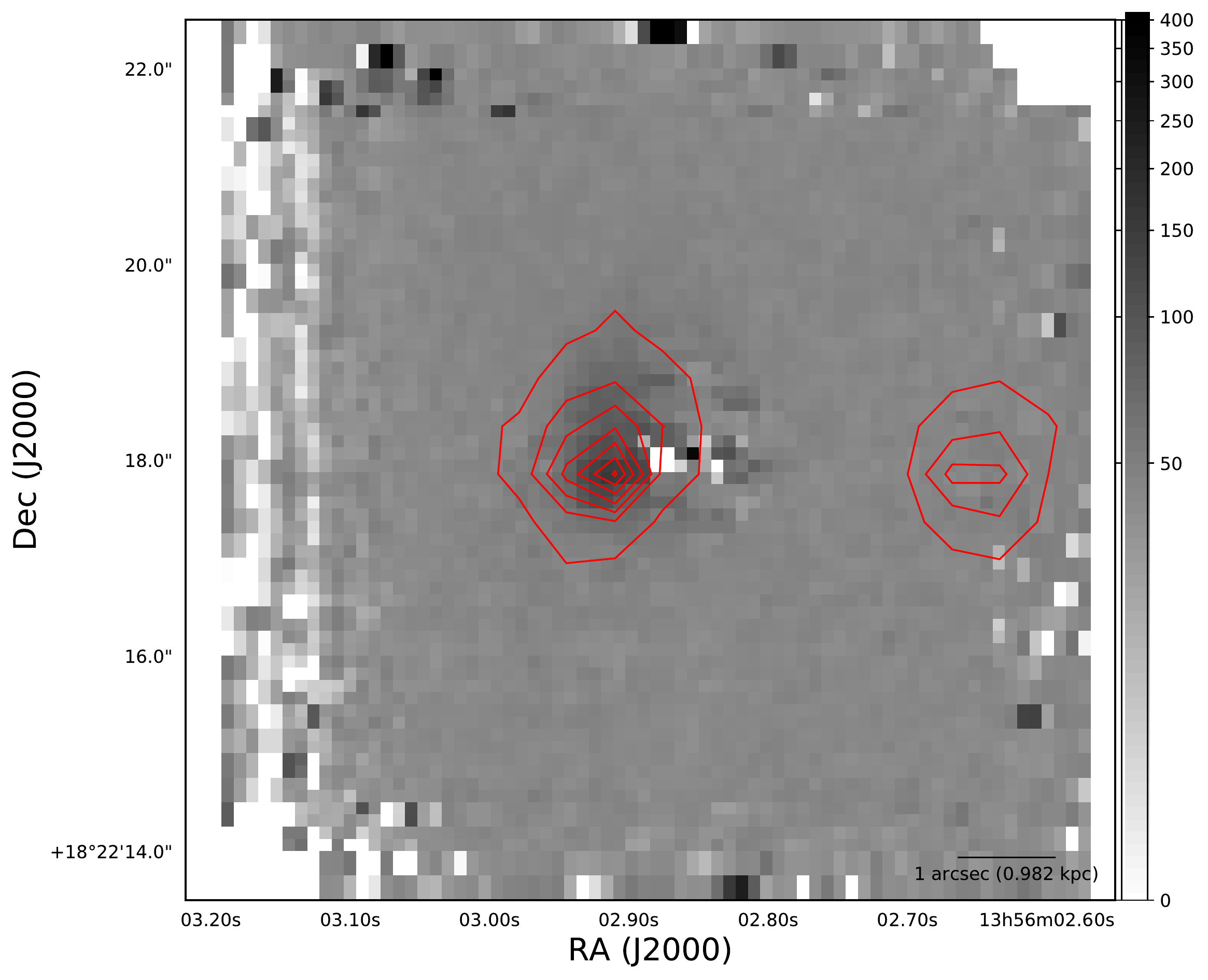}}
\end{center}
\vspace{-0.1in}
\caption{{\it Left panel:} Emission map for the [Si VI] emission line at 1.962 $\mu$m in the rest frame. In this case both nuclei are clearly detected.
{\it Right panel:} Image of the Hydrogen Br$\gamma$ emission at 2.1655 $\mu$m in the rest frame. The Br$\gamma$ appears to be significantly
fainter and hence only the Mrk\,463E nucleus is detected and at a much lower significance. Red contours in both cases show the hard X-ray emission 
detected by Chandra.\label{img_siVI_brgamma}}
\end{figure*}

The near-IR data also map other emission lines such as [Si VI]  and Hydrogen Br$\gamma$ at rest-frame wavelengths of 1.962 and 2.1655 $\mu$m, 
respectively; these are both presented in Figure~\ref{img_siVI_brgamma}. Emission from [Si VI] is detected at high significance from Mrk 463E and weakly from 
Mrk 463W, thus further confirming the AGN nature of both nuclei. However, Br$\gamma$ is in contrast only weakly detected from Mrk 463E.

\subsection{CO and dust continuum}
\label{sec:morph_co}

\begin{figure}
\vspace{-0.1in}
\begin{center}
\hglue-0.0cm{\includegraphics[width=3.5in]{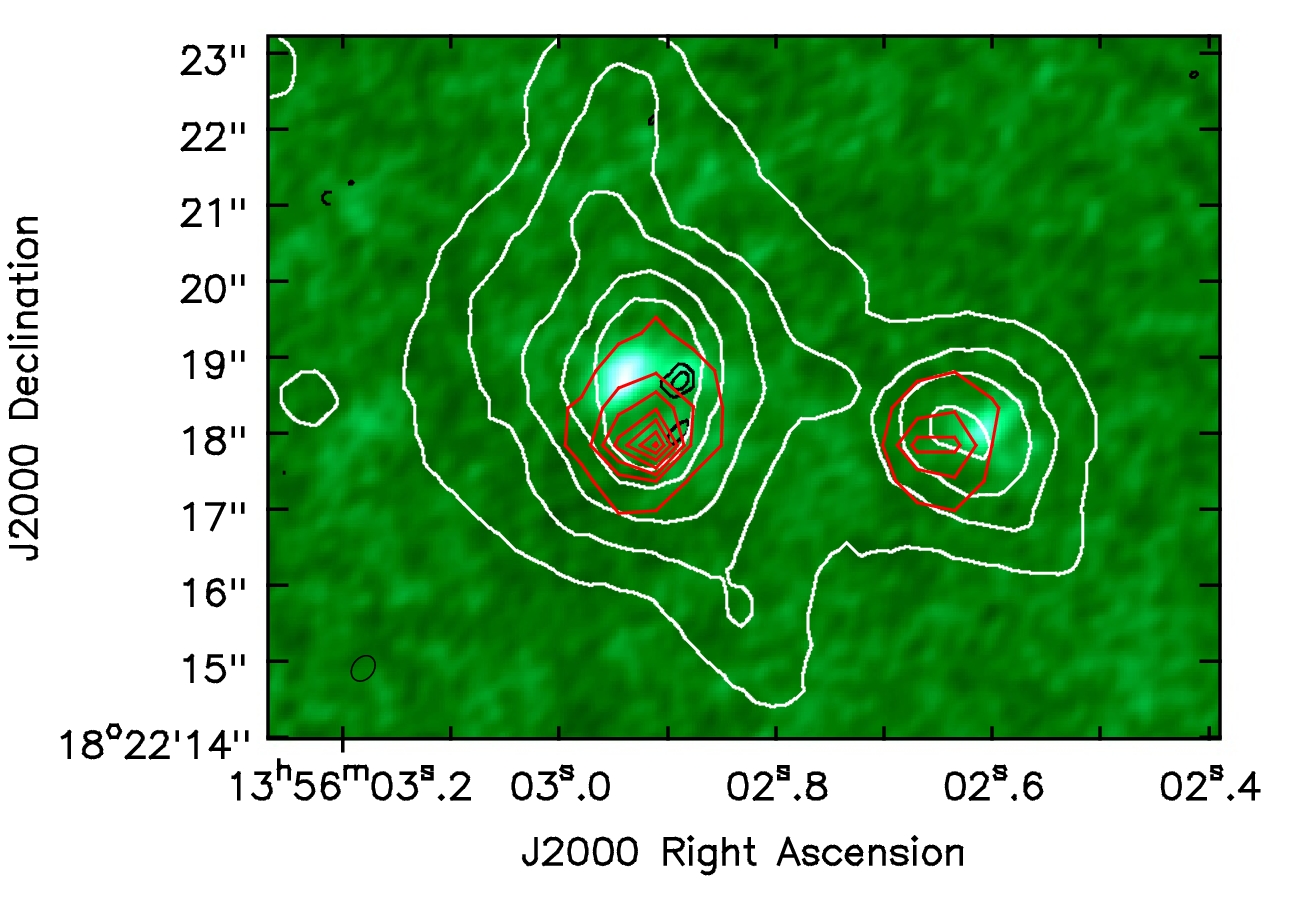}}
\end{center}
\vspace{-0.2in}
\caption{Mrk\,463 ALMA $^{12}$CO(2-1) emission integrated map. The {\it black contours} show the mm-continuum emission surrounding the $^{12}$CO(2-1) line, while 
the {\it red contours} present the {\it Chandra} hard X-ray, 2--8 keV, emission and the {\it white contours} present the background-subtracted $H\alpha$ flux obtained from 
MUSE. The ellipse in the bottom left shows the beam size for the $^{12}$CO(2-1) ALMA map. \label{img_co21_mom0}}
\end{figure}

With our ALMA Cycle 2 data we detect both $^{12}$CO(2-1) and the  mm continuum in the central region of the Mrk\,463 system. Fig.~\ref{img_co21_mom0} 
shows the velocity-integrated continuum-subtracted $^{12}$CO(2-1) emission map together with contours for the rest-frame 232 GHz continuum. As can be seen, 
the strongest CO(2-1) emission is found in a region to the Northeast of the Mrk\,463E nucleus, although significant emission is also located around each nucleus and 
in between them. Interestingly, only a small fraction of the CO(2-1) emission overlaps with the AGN location. 

Defining a $\sim$4$''$$\times$3$''$ region centered on the Mrk\,463E nucleus that encloses the majority of the flux, we obtain a total flux density
of 11.4$\pm$0.12 Jy km/s. In this case the error bars were estimated by measuring the RMS in nearby source-free regions of the same area. Previous
observations of this source at lower resolution, $\sim$2$''$, yielded total unresolved fluxes of 7.2$\pm$0.8 Jy km/s \citep{alloin92} and 6.8$\pm$0.9 Jy km/s 
\citep{evans02} for the $^{12}$CO(1-0) transition. Following \citet{solomon05}, we assume that the CO emission is thermalized and optically thick so that the intrinsic 
line luminosity is independent of $J$ and hence we expect the $^{12}$CO(2-1) flux density to be 4$\times$ higher than $^{12}$CO(1-0). Therefore, our high-resolution 
ALMA observations resolve out $\sim$50\% of the CO emission for the Mrk\,463E nucleus extended over scales of $\sim$4$''$. Using a region of the same area 
centered on the Mrk\,463W nucleus, we derive a line flux density of 4.5$\pm$0.12 Jy km/s. This is consistent with the estimates of \citet{evans02}, who also found a 
flux ratio of $\sim$3:1 between the Mrk\,463E and Mrk\,463W nuclei, thus suggesting that the extended component encompasses both nuclei, expected given the 
observed nuclear separation.

Following the prescription described in Sections 2.1 and 2.2 of \citet{solomon05}, and in particular their equation 3:
\[
L'_{CO}=3.25\times 10^7 S_{CO}\Delta v \nu_{obs}^{-2}D_L^2 (1+z)^3,
\]
we calculate a CO(2-1) luminosity for the emission surrounding the Mrk\,463E nucleus of $L'_{CO}$(2-1)=3.173$\times$10$^{8}$ K km s$^{-1}$pc$^2$. Then,
using the relation $M_{gas}$=M($H_2$)=$\alpha$$L'_{CO}$, and $\alpha$=4.3M$_\sun$(K km s$^{-1}$pc$^2$)$^{-1}$ \citep{bolatto13}, we obtain a total gas mass
in this region of 1.36$\times$10$^9$M$_\sun$. Similarly, for the material surrounding the Mrk\,463W nucleus we obtain $L'_{CO}$(2-1)=1.253$\times$10$^{8}$ K km s
$^{-1}$pc$^2$ and $M_{gas}$=5.39$\times$10$^8$M$_\sun$. These differ from the gas masses derived by \citet{evans02} primarily due to the different values 
of $\alpha$ assumed. Indeed, \citet{evans02} used a value of $\alpha$=1.5M$_\sun$(K km s$^{-1}$pc$^2$)$^{-1}$ for this source and reported a total
molecular gas mass of 10$^9$M$_\sun$, while assuming their $\alpha$ value we obtain $\sim$7$\times$10$^8$M$_\sun$.

\section{Kinematics}
\label{sec:kine}

The multi-wavelength IFU data for Mrk\,463 can be used to trace both absorption and emission lines and measure the 
kinematics of the system.

\subsection{Optical Emission Lines}
\label{sec:kine_opt_lines}

\begin{figure*}
\plottwo{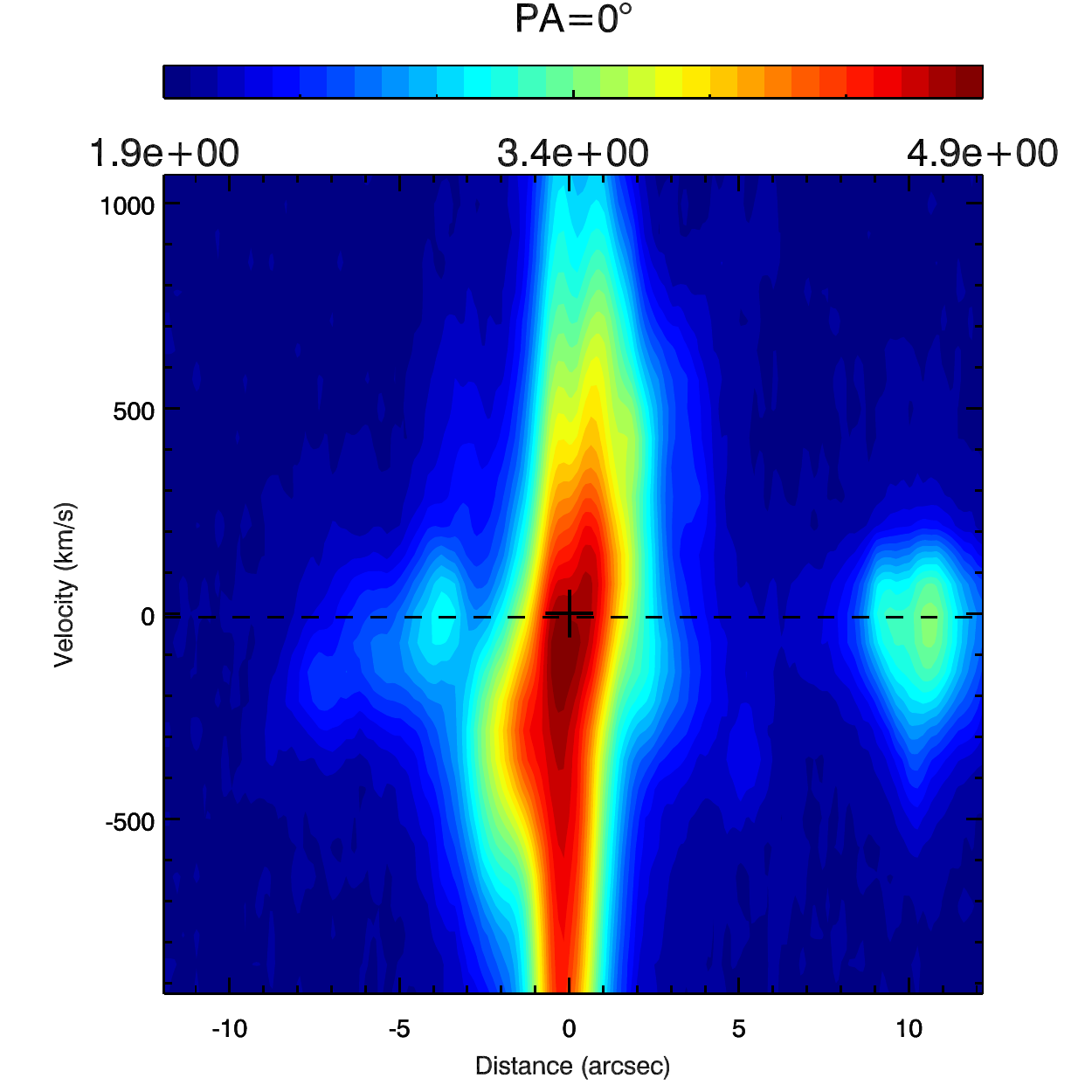}{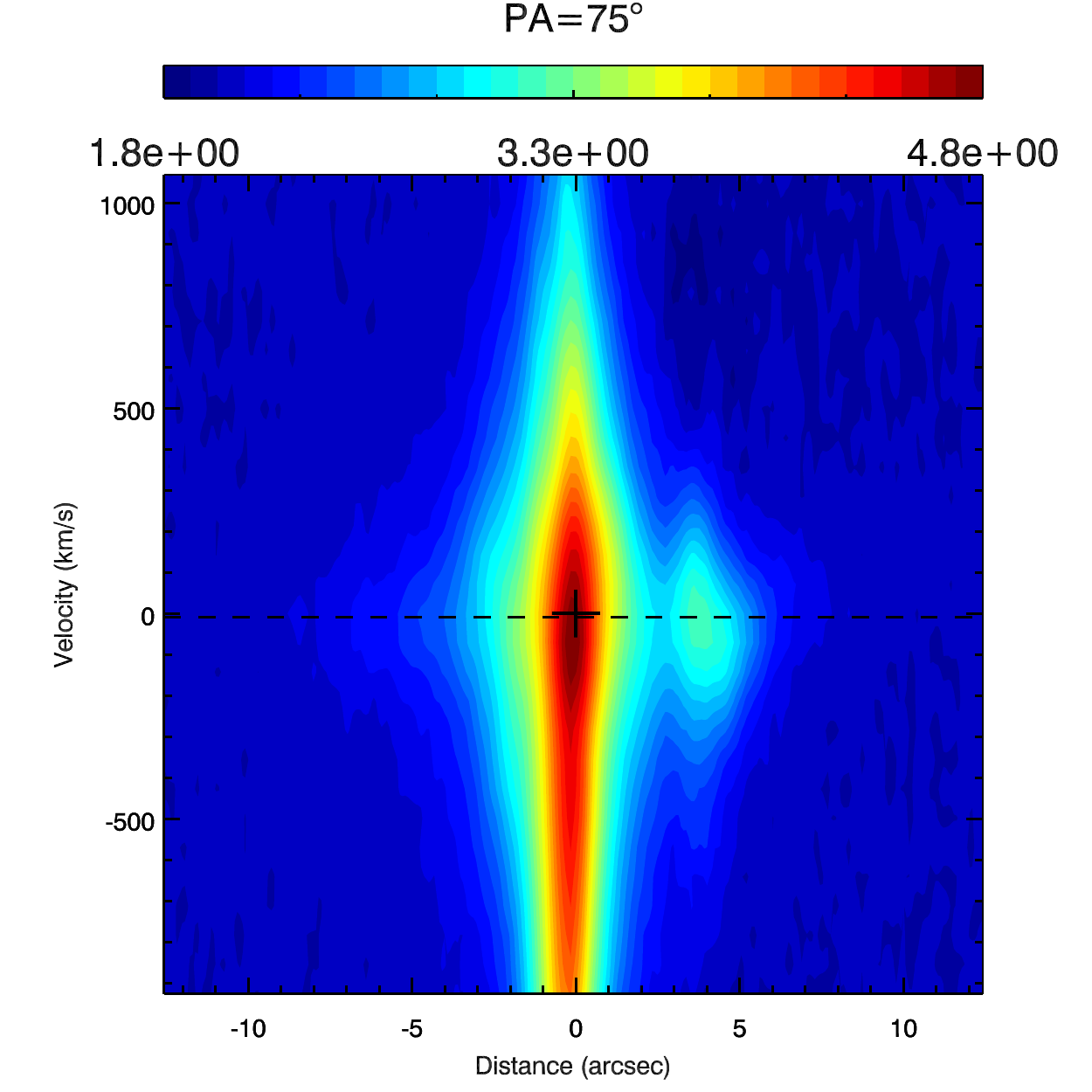}
\caption{Position velocity (p-v) diagram for the [O{\sc iii}] 5007\AA~ emission line centered on the East nucleus. The {\it left panel} shows the emission along
a north-south (PA=0$^\circ$) slit, while the {\it right panel} uses a PA of 75$^\circ$. The north-south slits falls directly on the position of the biconical outflow discussed
below, which can be clearly visible, while no significant structures are visible on the 75$^\circ$ direction. About 10$''$ away from the nucleus on the PA=0$^\circ$ diagram
we can see the southern emission region, while on the PA=75$^\circ$ slit we can see part of the western nucleus at $\sim$4$''$. \label{Mrk463_OIII_pv}}
\end{figure*}

Using the VLT/MUSE data, we analyze the kinematics from two strong and well-isolated emission lines: $H\beta$ 4861\AA~and [O{\sc iii}] 5007\AA. 
Figure~\ref{Mrk463_OIII_pv} shows the position-velocity (p-v) diagram centered on the Mrk463-E nucleus with position angles (PAs) of 0$^\circ$ (i.e., north-south) 
and 75$^\circ$ using the [O{\sc iii}] 5007\AA~line. In this case we assume a reference velocity of 15226 km/s, which is 130 km/s higher than the systemic
heliocentric velocity of 15096 km/s corresponding to the redshift of 0.050355 reported by \citet{falco99} which we used for the rest of the analysis. This was done
to match the center of the potential additional stellar disk to the west of Mrk 463W discussed in \S4.2. A clear velocity gradient can be seen on the north-south slit, as 
discussed below. This structure is highly localized spatially, as it is not present on the PA=75$^\circ$ diagram. Visible on the PA=0$^\circ$ slit is the southern emission 
line region at $\sim$10$''$. As can be seen, this region is roughly at the systemic velocity. Similarly, part of the emission associated with the Mrk463W nucleus can be 
seen on the PA=75$^\circ$ diagram, at $\sim$4$''$ away from the eastern one.

\begin{figure*}
\plottwo{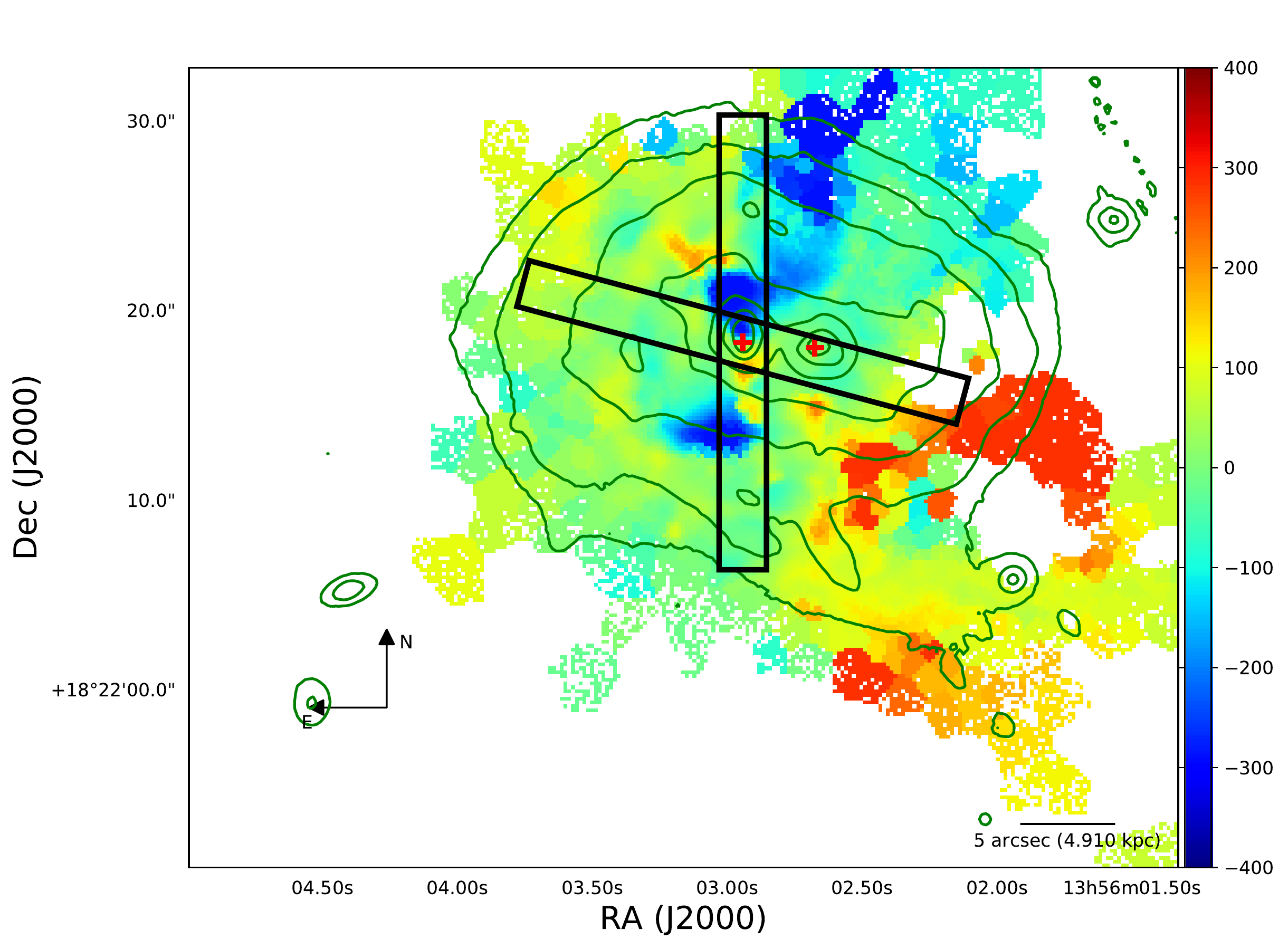}{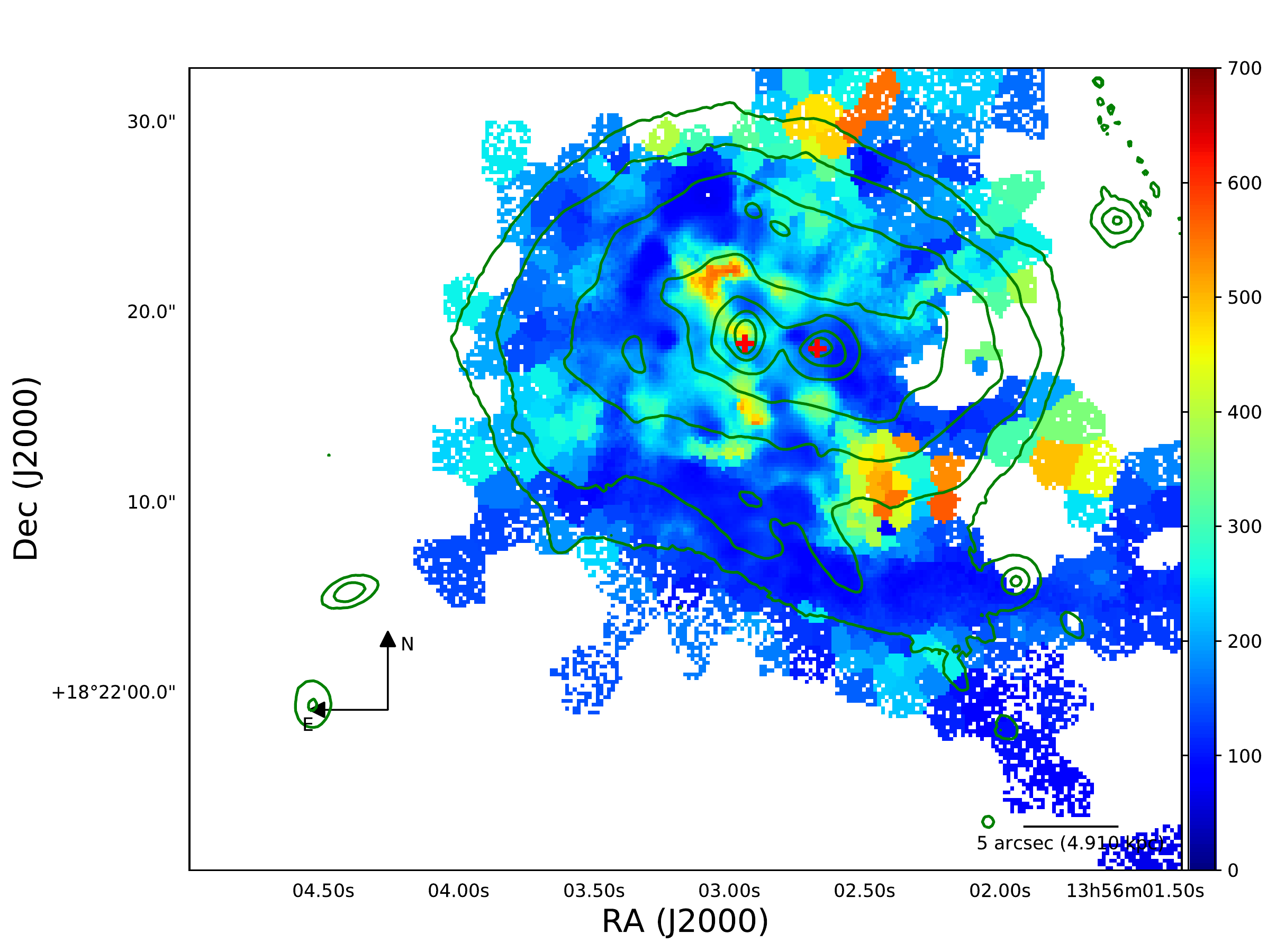}
\caption{{\it Left panel:} Velocity map for the [O{\sc iii}] 5007\AA~line, measure in km/s relative to the systemic velocity at the redshift of the system ($z$=0.050355). 
Contours and symbols are the same as in Fig.~\ref{img_Hbeta_OIII}. Black rectangles show the location of the slits used for the p-v diagrams presented in
Figure~\ref{Mrk463_OIII_pv}.{\it Right panel:} Velocity dispersion map for the [O{\sc iii}] 5007\AA~line. Contours are the same as in the left 
panel. \label{img_OIII_velmap}}
\end{figure*}

Figure~\ref{img_OIII_velmap} shows the velocity and velocity dispersion maps for the [O{\sc iii}] 5007\AA~emission line. The velocity gradient associated with the
eastern nucleus that was visible on the p-v diagram presented in Fig.~\ref{Mrk463_OIII_pv} is present here as well. As first suggested by \citet{hutchings89} and later 
discussed by \citet{chatzichristou95}, this structure is consistent with the presence of a two-sided ionization cone, which also aligns with the extended radio emission 
detected from this nucleus. While a detailed modeling of this biconical outflow is beyond the scope of this paper, we can use the observed emission line 
features in this region to derive physical properties of the ionized gas outflow. As presented by \citet{muller-sanchez11}, the mass outflow rate is given by

\[
\dot{M}_{\textnormal{out}}=2m_p n_ev_{max}A f,
\]
\noindent where $m_p$ is the proton mass, $n_e$ is the density of the ionized gas, $A$ is the lateral surface area of one cone of the outflowing region, and $f$ is the 
filling factor of the ionized gas. The factor of 2 accounts for the two sides of the ionized cone. The value of $n_e$ can be estimated from the observed ratio 
of the [SII] 6716/6731\AA~emission lines, as described by \citet{osterbrock06}. We find that on average the [SII] ratio in the region of the biconical 
outflow has values $\sim$0.7. According to the prescription of \citet{osterbrock06} and assuming a temperature of $\sim$10,000K, we estimate that $n_e$ should 
be in the range from 1000 to 5000 cm$^{-3}$ and adopt a value of 3000 cm$^{-3}$. For the filling factor, $f$, which cannot be derived directly from observations, a range 
of 0.01$<$$f$$<$0.1 is commonly assumed \citep[e.g.,][]{storchi-bergmann10}. However, our assumed $n_e$ value is much higher than in those cases, and considering the
$n_e$$\propto$$f^{-0.5}$ relation proposed by \citet{oliva97} we assume a value of $f$=0.001, which is 10 times larger than the value used by \citet{nevin18}. 
We consider a maximal velocity of $v_{max}=$350km/s, as measured on the [O{\sc iii}] velocity map. The value of $A$ is obtained considering that the physical 
distance from the nucleus to the location of the maximal velocity is $\simeq$3.3 kpc and that the cone radius at that position is $\simeq$0.9 kpc thus yielding
A=9.62$\times$10$^6$pc$^2$. Then, using the expression described above we obtain that $\dot{M}_{\textnormal{out}}$=512($f$/0.001) $M_\odot$/yr. This value 
is on the high side of the range of mass outflow rates found previously in systems hosting biconical outflows \citep{muller-sanchez11,muller-sanchez16,nevin18} in 
moderate-luminosity AGN and similar to those found in high redshift quasars \citep{brusa15} and ULIRGs \citep[e.g.][]{veilleux05}.

In addition, we also detect a rapidly outflowing region to the north-west of the Mrk\, 463E nucleus, reaching speeds of up to $\sim$-600 km/s. 
This region was already identified as \#4 in Figure~\ref{img_line_ratios}. As shown in Figure~\ref{bpt}, this region has the highest value of the [O{\sc iii}]/H$\beta$ 
ratio and is clearly located in the AGN region, which confirms that this outflowing material is being ionized by the AGN, most likely in Mrk\,463E given its spatial location 
and the luminosity of its nucleus. Furthermore, this outflowing region appears to be spatially connected to the biconical outflow. The velocity dispersion map for the 
[O{\sc iii}] 5007\AA~line shows a broad emission line component in the central region of Mrk\, 463E. Both the north and south edges of the biconical outflows also
show relatively high velocity dispersions, reaching up to $\sim$600 km/s, and hence suggest the presence of shocks in the boundary of the outflowing regions.
The remaining regions are characterized by relatively low velocity dispersions, up to $\sim$200 km/s, close to the spectral resolution of VLT/MUSE.

\begin{figure*}
\plottwo{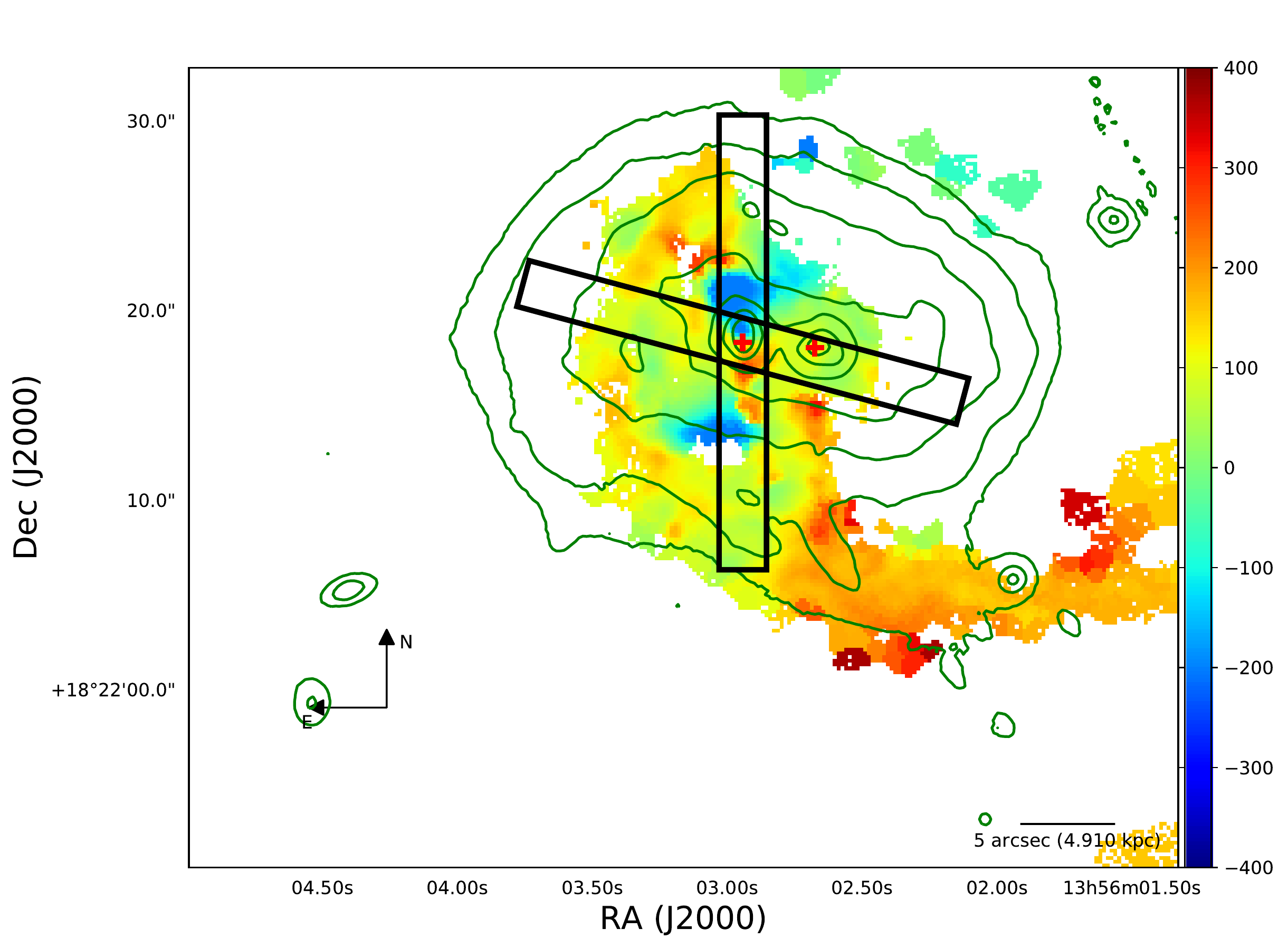}{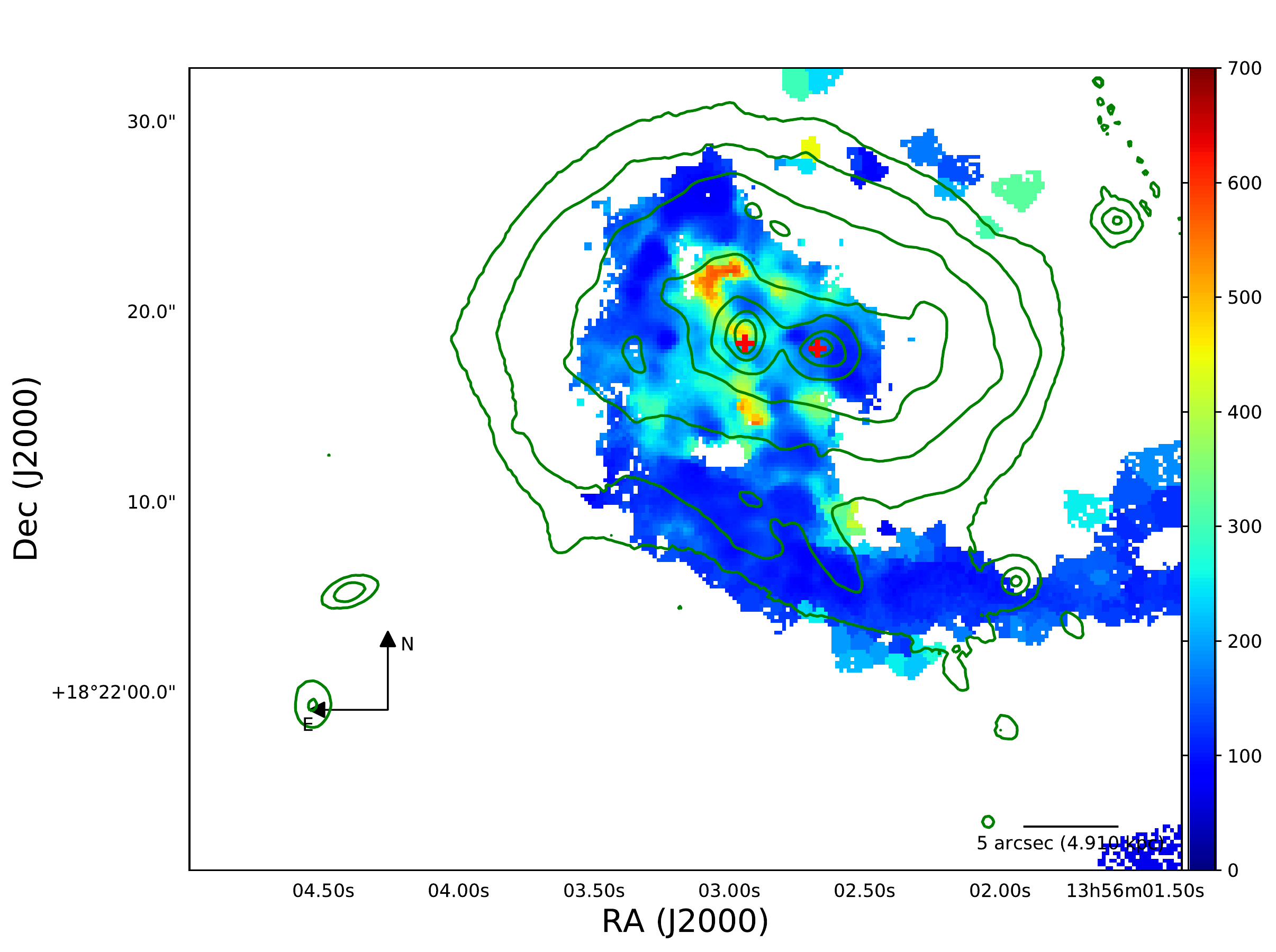}
\caption{{\it Left panel:} $H\beta$ 4861\AA~velocity map in km/s relative to the systemic velocity. Contours 
show the $H\beta$ 4861\AA~ flux, as measured from our Gaussian fitting of the line, presented in Figure~\ref{img_Hbeta_OIII}. 
Black rectangles show the location of the slits used for the p-v diagrams presented in
Figure~\ref{Mrk463_OIII_pv}{\it Right panel:} $H\beta$  4861\AA~velocity dispersion map in km/s. Contours are the same as in the left panel.\label{img_Hbeta_velmap}}
\end{figure*}

The velocity and velocity dispersion maps, as measured from the $H\beta$ 4861\AA~emission line, are shown in Figure~\ref{img_Hbeta_velmap}.
As for the [O{\sc iii}] 5007\AA~line, a very clear gradient, ranging from -350 km/s in the north to $\sim$150 km/s to the south, centered on the
Mrk\,463E nucleus is clearly visible. This gradient is also associated with the biconical outflow presented above. This map does not include the 
absorption line region located to the west of the Mrk463W nucleus, which will be studied in detail in the following subsection. Both nuclei appear to 
be consistent with the systemic velocity, or with very small offsets. Similarly, the southern emission line region is at the systemic velocity and has 
no discernible structure. However, there is a clear tail of atomic gas emission traced by the $H\beta$ 4861\AA~transition, which includes the southern
emission line region visible in the [O{\sc iii}] map, presenting a velocity gradient ranging from $\sim$100 km/s to $>$200 km/s. This appears to reflect 
the velocity gradient along a tidal tail from the ongoing dynamical interaction and is consistent with the velocity gradient seen in HI observations of 
other systems  \citep[e.g.,][]{hibbard96}. 

The velocity dispersion profile for the $H\beta$ 4861\AA~line reveals the presence of a relatively broad component with a FWHM of up to $\sim$700~km/s in the 
nuclear region Mrk\,463E, consistent with the location of the hard X-ray emission reported by \citet{bianchi08}. The arc-like structure to the north-east of
the Mrk\,463E, which also shows high velocity dispersion values, is explained by the presence of two superimposing velocity components, one associated
to the edge of the biconical outflow reaching velocities of $\sim$400 km/s and the underlying galaxy at the systemic velocity. The rest of the system presents 
only narrow components with a width of $\sim$200 km/s or less, consistent being unresolved at the MUSE spectral resolution.

\subsection{Stellar kinematics}
\label{sec:kine_stellar}

\begin{figure*}
\plottwo{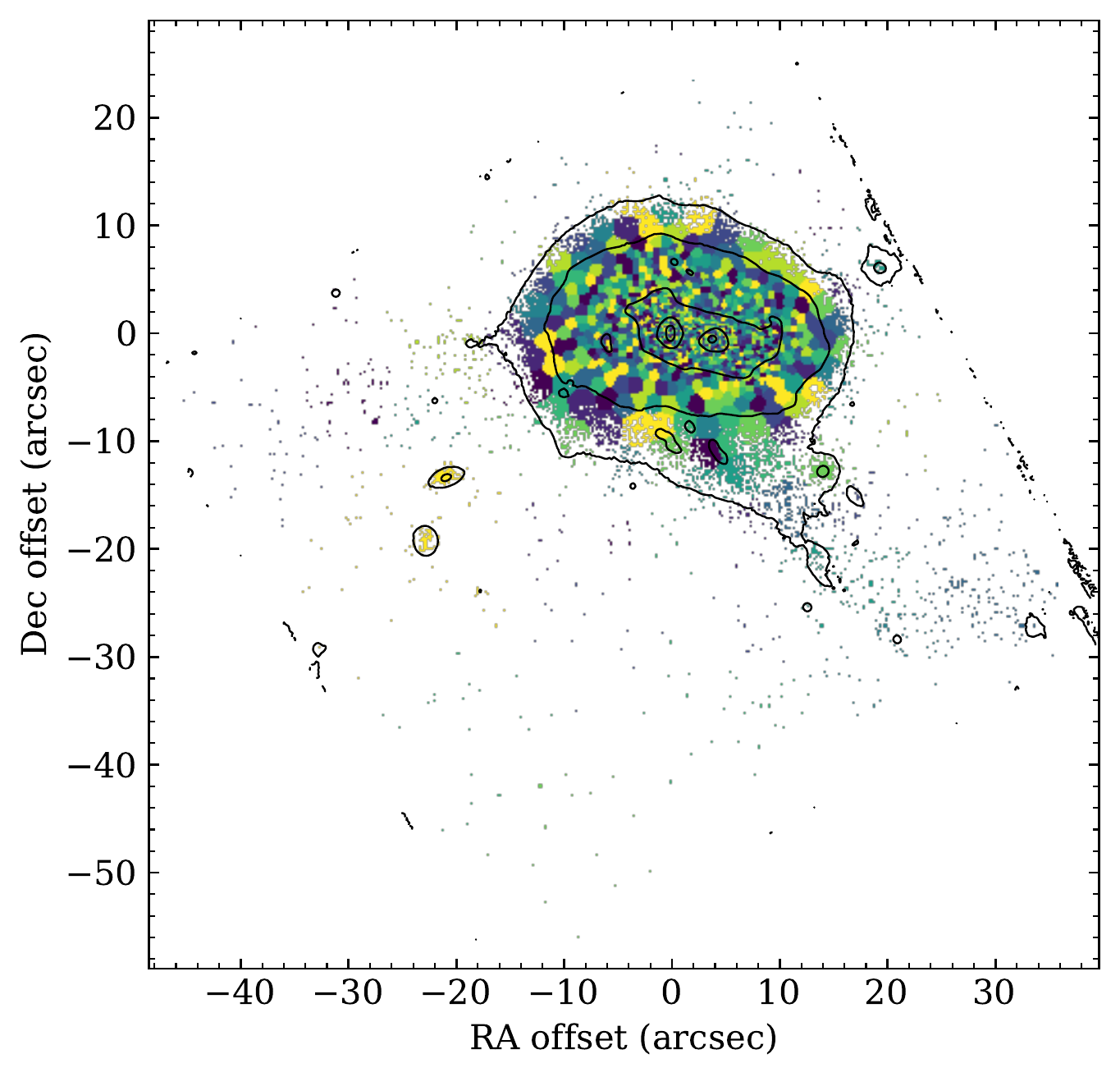}{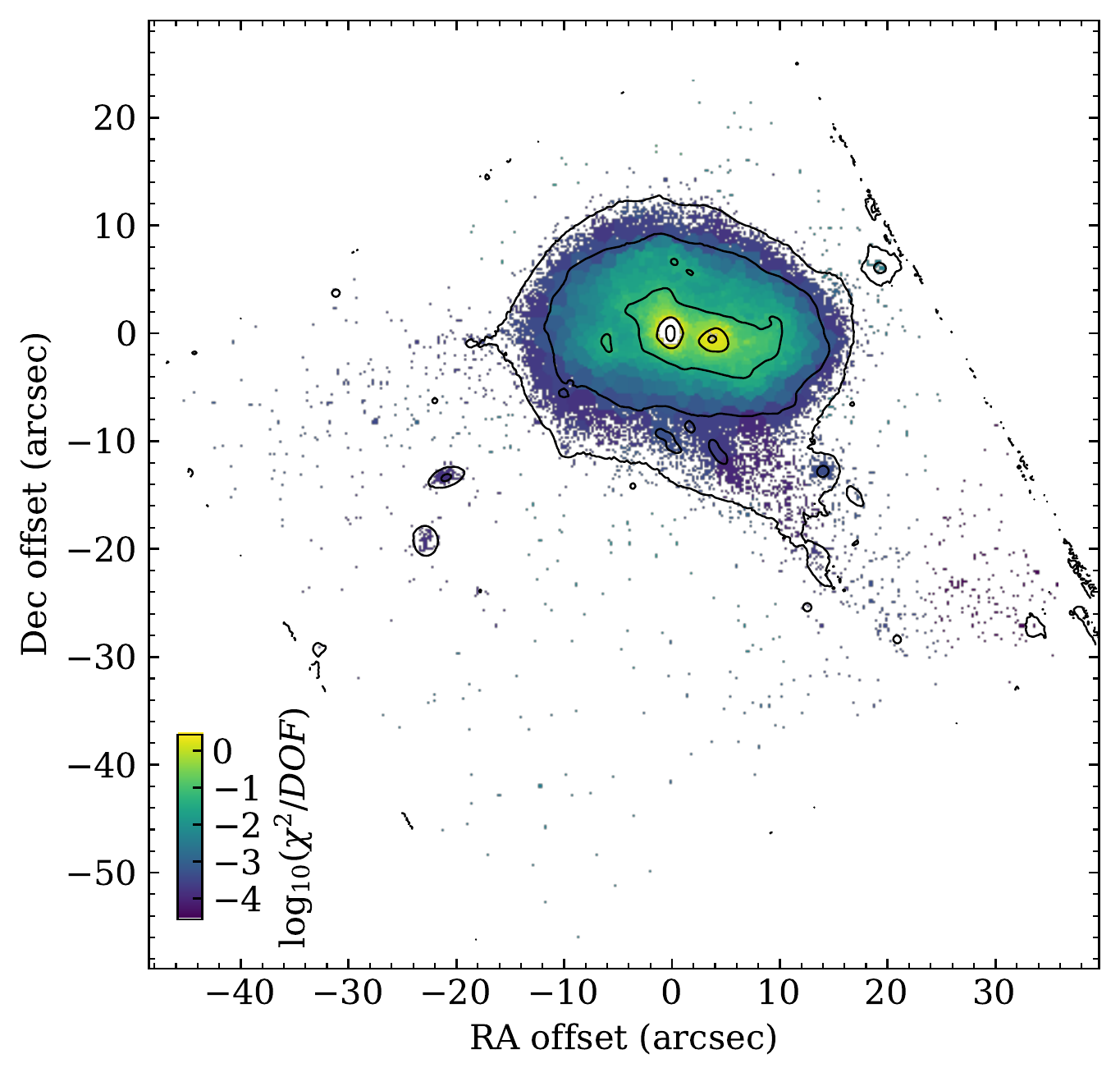}
\caption{{\it Left panel:} Voronoi tessellation binning of the Mrk\,463 VLT/MUSE data, requiring a minimum SNR of 30.  The black contours were obtained from the white light 
(integrated) VLT/MUSE image, presented in Figure~\ref{MUSE_cont}. The lowest contour is at 2$\times$10$^{-20}$ erg s$^{-1}$cm$^{-2}$\AA$^{-1}$, while the contours 
increase as 2$^n$. The coordinates are given relative to the position of the Mrk\,463E nucleus at 13:56:02.92,18:22:18.1. {\it Right panel:} Reduced $\chi^2$ map for the 
stellar population fits. Contours and coordinates are the same as in the left panel. Fits were not performed in the nuclear regions, as in these areas the emission is strongly 
dominated by the AGN.\label{stellar_bins_chi2}}
\end{figure*}

We can further use the VLT/MUSE data cube to measure the kinematic properties of the stars in the Mrk\,463 system. In order to achieve
this, we performed a new Voronoi tessellation \citep{cappellari03}, this time aimed to reach a minimum signal-to-noise ratio of 30 per bin in 
the optical continuum near the Mg~b feature at $\sim$5200\AA~in the rest frame. The resulting bin map is shown in Figure~\ref{stellar_bins_chi2}. Then, for each bin, we simultaneously masked the 
most prominent emission lines and fitted the optical continuum and absorption lines using the Penalized Pixel-Fitting (pPXF) method presented by \citet{cappellari04}. The 
optical spectrum at each Voronoi bin was fitted using stellar templates from the MILES library \citep{falcon-barroso11} which were then used to derive properties such as the stellar velocity and velocity dispersion, $h_3$ and $h_4$ Hermite polynomial coefficients, etc. A more detailed description of this procedure is given by \citet{cappellari17}. In Figure~\ref{stellar_bins_chi2} we show the reduced $\chi^2$ for each Voronoi bin. Fits were not obtained for the central regions of each galaxy, as 
the optical spectra in these are dominated by AGN light. 

\begin{figure*}
\plottwo{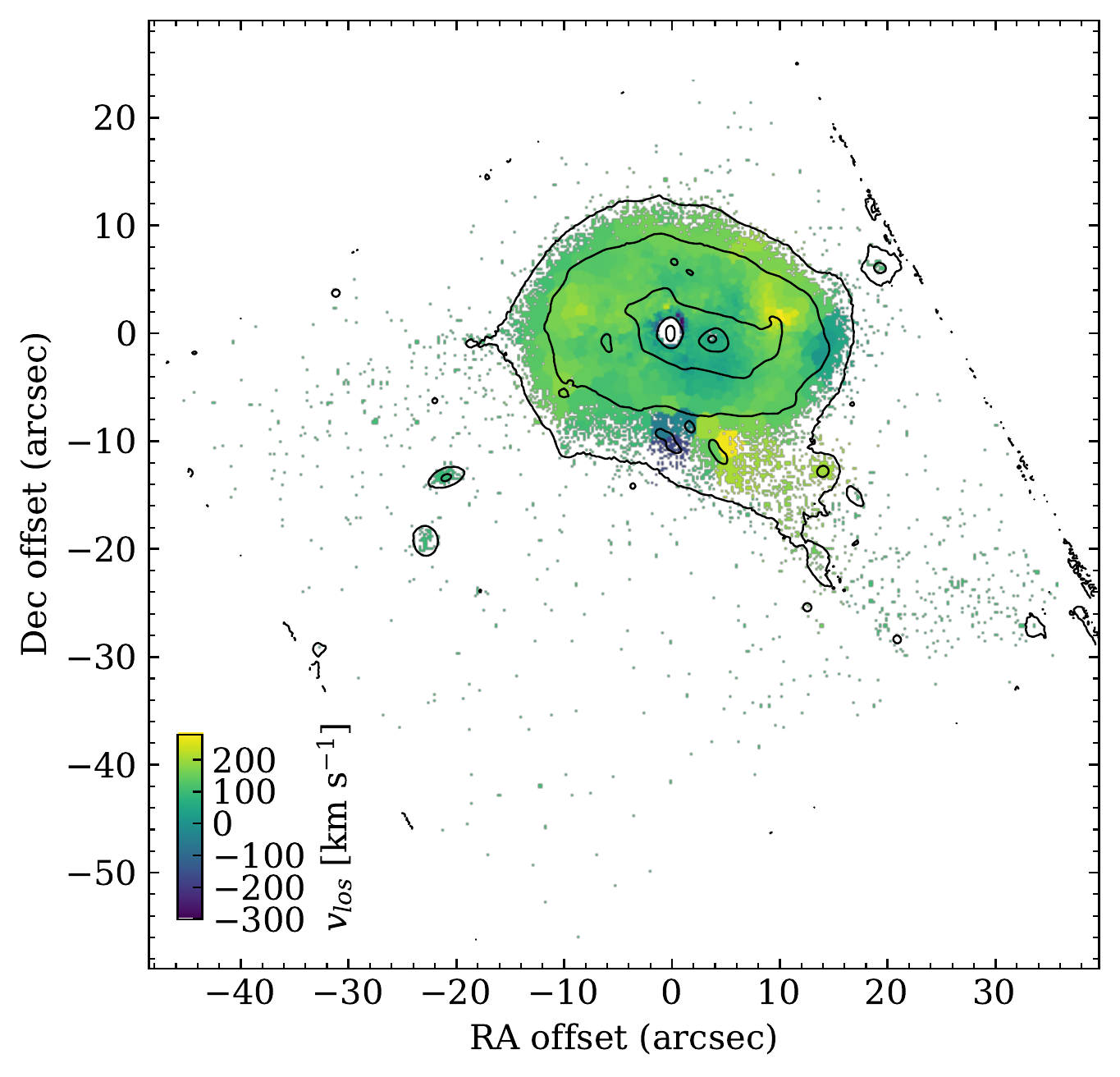}{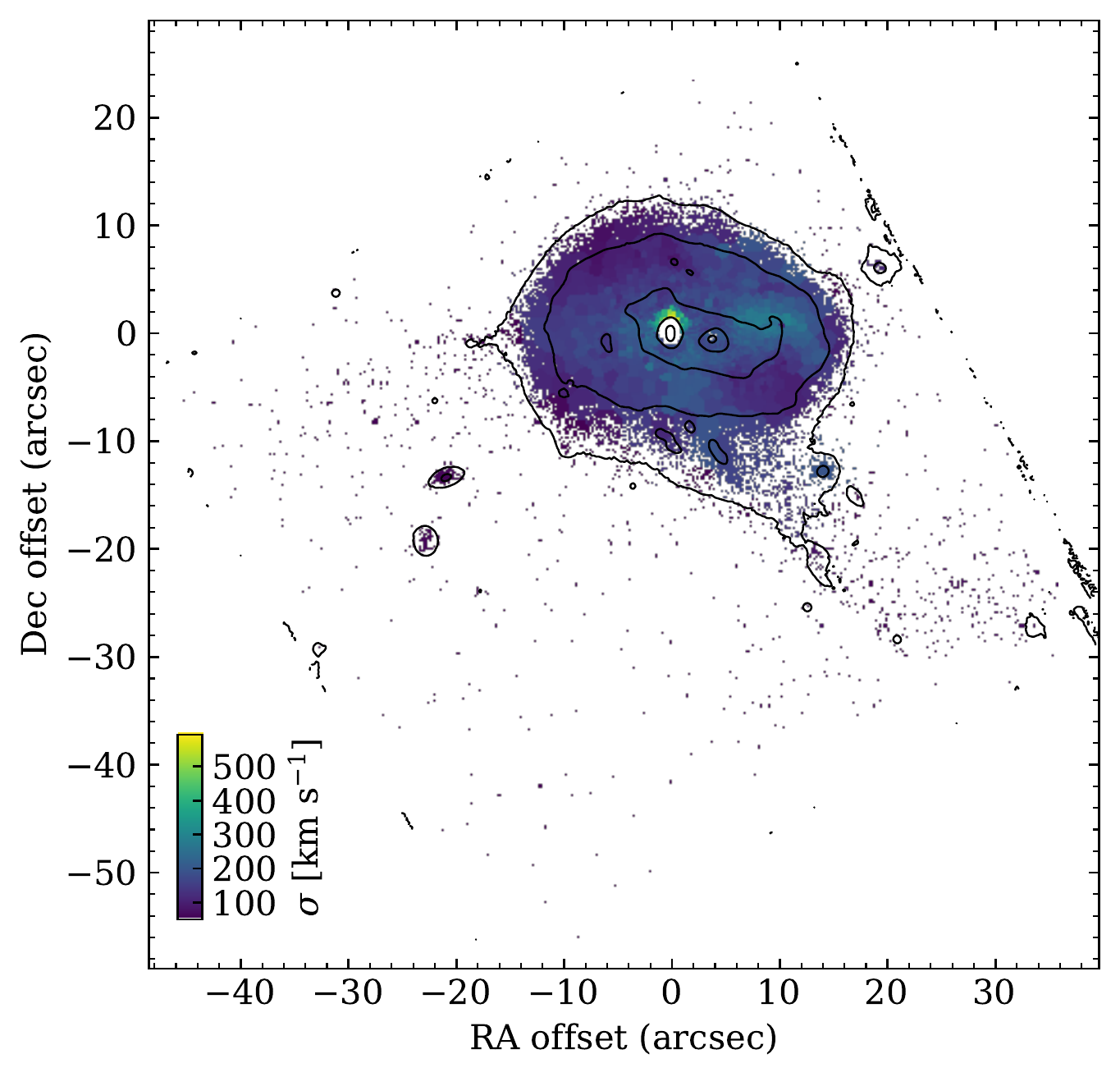}
\caption{Stellar velocity ({\it left panel}) and velocity dispersion ({\it right}) maps for the Mrk\,463 system, in km s$^{-1}$.  Contours and coordinates are the same as in Fig.~\ref{stellar_bins_chi2}. \label{stellar_vel}}
\end{figure*}

By performing this spectral fitting, in Figure~\ref{stellar_vel} we present the stellar velocity and velocity dispersion maps. While the stellar velocity map appears rather
flat, some structures are visible. In particular, there is a clear region to the northwest of the Mrk\,463W nucleus presenting a relative line-of-sight velocity of $\sim$200~km/s. 
In addition, a negative velocity region can be observed to the south of the Mrk\,463W nucleus. These features could potentially be related to the ongoing major merger, such
as tidal plumes from a second (or latter) close pass of the nuclei. Interestingly, this structure overlaps spatially with a region dominated by H$\beta$ in absorption, as 
can be seen in Figure~\ref{img_Hbeta_OIII}. This, together with the fact that this region does not have a counterpart on the emission line velocity maps presented in 
Figures \ref{img_Hbeta_velmap} and \ref{img_OIII_velmap}, strongly suggests that this region is mostly devoid of ionized gas. While very faint, this structure can be
seen as an extra component in the optical continuum, as shown in Figure~\ref{MUSE_cont}.

The stellar population fit simultaneously constrains the line-of-sight stellar velocity dispersion, which is presented on the right panel of Figure~\ref{stellar_vel}. As for the 
velocity map, the velocity dispersion in the Mrk\,463 system appears to be rather flat and featureless.  Again, the region to the northwest of the Mrk\,463W nucleus presents 
an elevated value for the stellar velocity dispersion of $\sim$400 km s$^{-1}$ while the average for the system is $\sim$100 km s$^{-1}$. This broadening in the 
line-of-sight velocity dispersion may be evidence of multiple velocity components at this location, perhaps due to the superposition of a secondary tidal tail against a 
potential stellar disk. 

\begin{figure*}
\plottwo{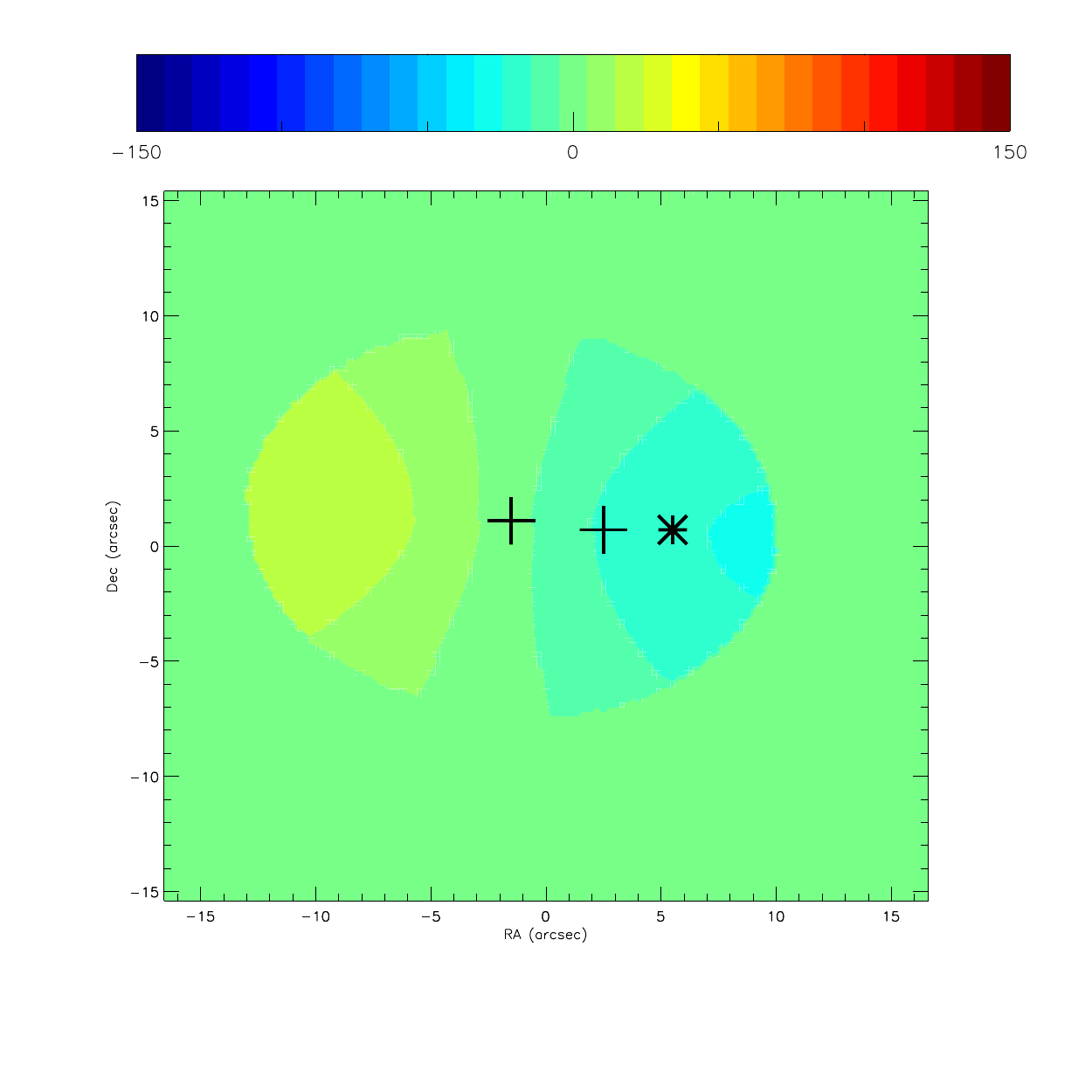}{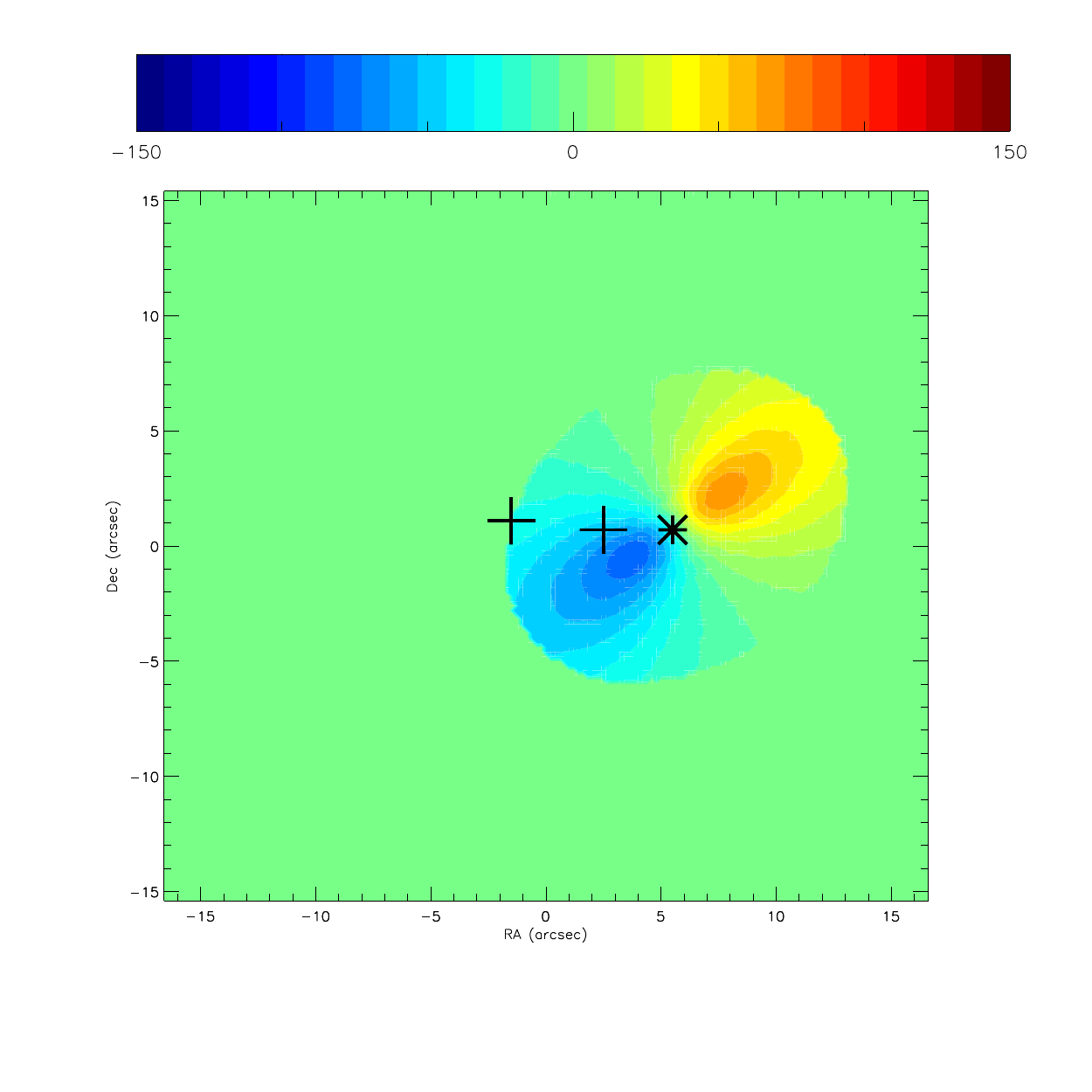}
\caption{Keplerian disk models used to fit the stellar kinematic maps shown in Fig.~\ref{stellar_vel}. Crosses show the positions of the two nuclei. Two disks were fitted, 
one centered on the Mrk\,463E nucleus and a PA of 75$^\circ$ (left panel) and a second one centered in the position marked by the asterisk and a PA of 125$^\circ$ 
(right panel).\label{stellar_disk}}
\end{figure*}

In order to explore the existence of potential disk(s) revealed by kinematical structures, we attempt to fit the stellar kinematics with disk models. Specifically,
we use the Keplerian disk models presented by \citet{bertola98}. We use two disks, one centered on the Mrk\,463E nucleus and the second one at the center of
the velocity gradient found to the west of Mrk\,463W. Spatial profiles of the two models are presented in Figure~\ref{stellar_disk}. The models fit the data reasonably well,
leaving only small velocity residuals. Hence, it is possible to explain the Mrk 463 system as a combination of a bulge and a faint extended disk associated with
Mrk\,463E, together with a bulge centered on Mrk\,463W and a gas-stripped stellar disk to the west of Mrk\,463W. There does not seem to be any connection 
between the ionized gas outflows discussed in section~\ref{sec:kine_opt_lines} and the potential stellar disks presented here. As discussed in \S\ref{sec:kine_opt_lines}, 
the outflows in this system appear to be mostly located in the north-south direction and originate from the Mrk 463E nucleus. In contrast, the Mrk 463E stellar disk 
is oriented with a PA of 75 degrees, i.e., mostly east-west. Similarly, the presumed stellar disk to the west of Mrk 463W does not appear to be connected
in any obvious way with any ionized gas outflow.

As described above, we would like to further emphasize that it is also possible to understand the observed structures in the stellar kinematics maps by 
the dynamical effects of the ongoing major galaxy mergers, which is also more natural as it does not require to assume an ad-hoc stellar disk offset by $\sim$3~kpc 
from the nucleus of the galaxy. Unfortunately, separating these two scenarios requires observations at higher spectral resolutions than those that can be provided by 
VLT/MUSE over relatively large angular scales in order to cover the whole system.

\subsection{CO and dust continuum}
\label{sec:kine_co}

\begin{figure}
\plotone{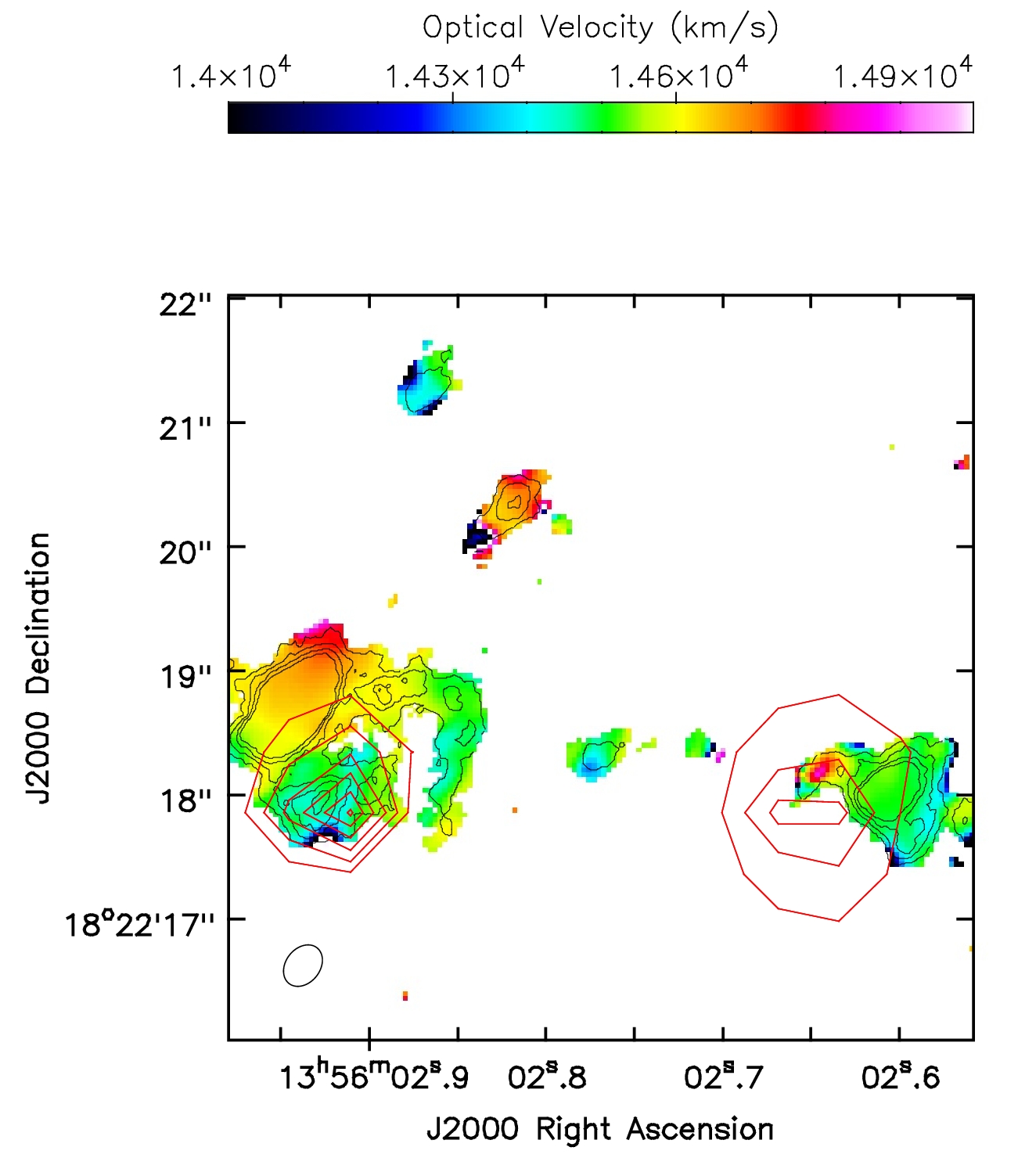}
\caption{$^{12}$CO(2-1) velocity map. Only regions with emission at $>$0.07 mJy/beam km/s are included in the map. {\it Red contours} correspond to
the {\it Chandra} hard X-ray, 2--8 keV, emission. \label{img_co21_velmap}}
\end{figure}

Thanks to the ALMA data, we can analyze the molecular gas kinematics in the nuclear regions of the Mrk\,463 system. Figure~\ref{img_co21_velmap}
shows the $^{12}$CO(2-1) velocity map. Interestingly, the molecular gas regions around each nucleus show very clear velocity gradients, 
with a range of $\sim$800 km/s in the case of the east region and $\sim$400 km/s next to the west region. Smaller gradients can be seen 
in a few other locations, including the one in between the two nuclei. 

The eastern emission region presents an incomplete ring-like or infalling spiral structure, possibly centered on the AGN, based on the X-ray position. While there is 
relatively little CO(2-1) emission detected directly on top of the nucleus, we can use the gas mass and velocity in the immediate vicinity of Mrk\,463E to obtain an 
estimate of the gas infall rate on scales of $\sim$300 pc around the SMBH. Considering only the material in the central ALMA beam, corresponding to $\sim$0.3$''$ or 
300 pc or a 150 pc radius, we obtain a total CO mass of $\sim$3.8$\times$10$^7$M$_\sun$. Now, assuming that this gas is falling at $\sim$200 km/s 
based on the CO(2-1) velocity measured on the nuclear ALMA beam and that it is at an average distance of 75 pc, this yields an influx rate of $\sim$104 M$_\sun$/yr. 
While just an approximation, this value is $>$3 orders of magnitudes larger than the accretion rate of $\sim$0.017 M$_\sun$/yr inferred from the X-ray luminosity of the 
Mrk\,463E nucleus reported by \citet{bianchi08}. This situation is very similar to what was found in the local AGN NGC1068 by \citet{muller-sanchez09}  and
later confirmed by \citet{garcia-burillo14} using high-resolution ALMA data, thus strongly suggesting that SMBH accretion is a highly inefficient process that 
requires the infalling material to lose most of its angular momentum.

\section{Discussion}

\subsection{Optical Extinction}\label{sec:opt_ext}

\begin{figure}
\vspace{-0.1in}
\begin{center}
\hglue-0.0cm{\includegraphics[width=3.5in]{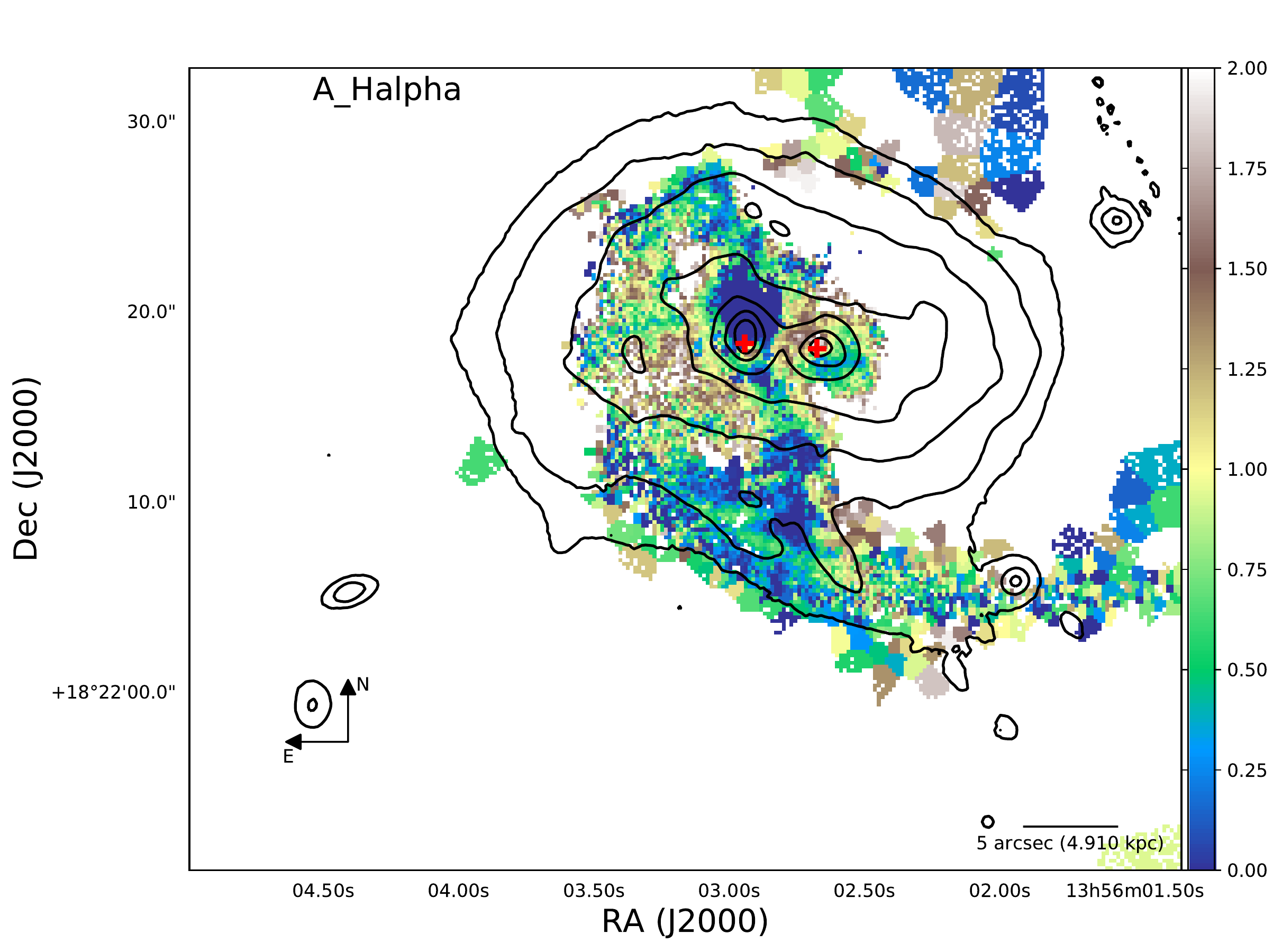}}
\end{center}
\vspace{-0.2in}
\caption{Optical extinction map for the Mrk 463 system derived from the $H\alpha$/H$\beta$ flux ratio, as described in the text. {\it Black contours} show 
the optical continuum flux, while the {\it red crosses} mark the location of the hard X-ray sources detected by Chandra. The map shows that while most of the
system is characterized by relatively low extinction levels, $A_{H\alpha}$$<$1, some specific regions, most notably in between the two nuclei, can reach
moderate absorption levels up to $A_{H\alpha}$$\sim$2.}
\label{img_AHalpha}
\end{figure}

Major galaxy mergers are known for potentially being subject to significant extinction, particularly at optical and UV wavelengths \citep[e.g.][]{trentham99,bekki00}.
Spectroscopically, the traditional way to estimate the amount of optical extinction is based on the observed ratio between the H$\alpha$ and H$\beta$ lines 
\citep[e.g.][]{caplan86,maiz-apellaniz04}. Taking advantage of the VLT/MUSE observations of the Mrk 463 system we can then produce a A$_{H\alpha}$ map. Following
the prescription of \citet{lee09}, we assume case B recombination with an intrinsic $H\alpha$/H$\beta$ ratio of 2.86, a Milky Way extinction curve and $R_V$=3.1 so that
A$_{H\alpha}$ is given by:
\[
A_{H\alpha}=5.91\log\left(\frac{f_{H\alpha}}{f_{H\beta}}\right)-2.70,
\]
where $f_{H\alpha}$ and $f_{H\beta}$ are the total emission line fluxes on the H$\alpha$ and H$\beta$ lines derived following the procedure described in 
section~\ref{sec:morph_optlines}. The resulting map is shown in Figure~\ref{img_AHalpha}. As can be seen, most of the system is only affected by relatively low
extinction levels, $A_{H\alpha}$$<$1. However, there are specific areas which appear subject to higher optical obscuration. These include some of
the regions where we previously identified high levels of star formation to the east of the Mrk 463E nucleus, and the area between the two nuclei, where 
we can see values of  $A_{H\alpha}$$\sim$1.5-2. While locally important, these relatively modest obscuration levels do not affect significantly the conclusions about the
morphologies and kinematics derived from the optical VLT/MUSE IFU maps presented here and are even less important for the near-IR
analysis shown in section~\ref{sec:morph_near-IR}.

\subsection{Energetics of the Emission Line Region}

As described in Section~\ref{sec:morph_optlines}, we identify an emission line region to the south of the system, consistent with AGN powering. This
region closely resembles the ``Hanny's Voorwerp'' structure associated with IC 2497 \citep{lintott09}, which was found serendipitously by visual inspection 
of the galaxy optical images carried out as part of the Galaxy Zoo project \citep{lintott08}. Hanny's Voorwerp appears to be ionized by the nearby AGN in IC~2497, 
which must have been at least 100 times more luminous $\sim100,000$~years earlier \citep{schawinski10a,keel12,sartori16}. For Mrk\,463, the southern region 
is $\sim$11 kpc away from the Mrk\,463E nucleus, which corresponds to $\sim$40,000 light years. Based on the evident alignment between this emission line
region and the ionization cone arising from the Mrk\,463E core, we can safely neglect the contribution from the Mrk\,463W AGN. Then, we can 
estimate the luminosity necessary in order to supply the ionizing flux required to explain the observed spectrum of the southern region.

In order to carry out this computation we selected a region in the center of the emission region enclosing most of its flux. We used as constraints the observed  
luminosities and their errors in the $H\alpha$, $H\beta$, [HeII]4686\AA, [O{\sc iii}]5007\AA, [NII]6583\AA, [SII]6717\AA~and [SII]6731\AA~lines, all measured from 
the VLT/MUSE optical IFU observations. We then performed radiative transfer simulations using the plasma ionization code CLOUDY version 13.04 \citep{ferland13}. We 
considered two different cases: (1) a matter-bounded scenario in which the main parameters are U, the dimensionless ionization parameter, n$_H$, the hydrogen 
volume density and N$_H$, the hydrogen column density, and (2) a radiation-bounded case, whose only parameters are U and n$_H$. In both cases we 
assumed that the input AGN spectrum was given by a broken power law of the form $L_\nu\propto$$\nu^\alpha$ with $\alpha$=-0.5 for E$<$13.6 eV, $\alpha$=-1.5 
in the range 13.6$<$E$<$0.5 keV and $\alpha$=-0.8 at E$>$0.5 keV, consistent with the observed spectrum of local AGN \citep[e.g.,][]{elvis94,kraemer09}. 

\begin{figure*}
\plotone{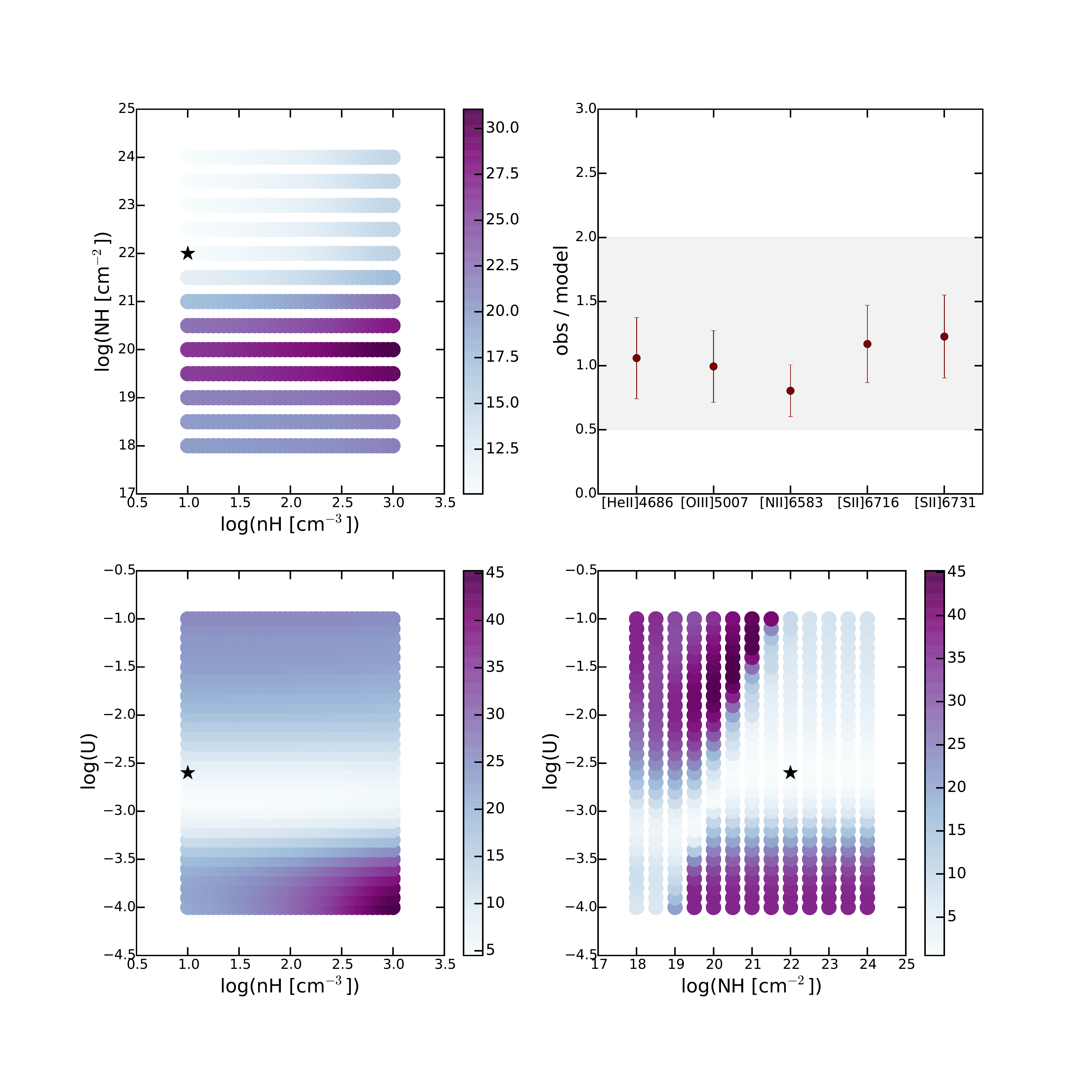}
\caption{Results of the CLOUDY simulation for a representative point inside the southern emission line region. The top left panel shows the resulting
column density, $N_H$, as a function of the volume density, $n_H$. Top right panel presents the ratio of observed to modeled flux for the best-fitting 
set of parameters and the corresponding error bars. The bottom left panel shows the values of reduced $\chi^2$ for the $\log$U and $\log$$n_H$ 
combinations while the bottom right panel presents the relation for $\log$U and $\log$$N_H$. The color scheme in all panels represents the 
reduced $\chi^2$, while the star shows the location of the best-fitting (minimum reduced $\chi^2$) solution.\label{cloudy_res}}
\end{figure*}

For each resolution element in the southern emission region we used the grid of simulations carried out using CLOUDY in order to find the best fit to the intrinsic 
emission line ratios, all computed with respect to the $H\beta$ line. From this solution we can obtain the corresponding value of Q, the number of ionizing photons 
emitted per unit of time, which is defined by

\begin{equation}
U=\frac{Q(H)}{4\pi r_0^2n(H)c},
\label{eq:lum}
\end{equation}

\noindent where $r_0$ is the distance between the AGN and the cloud. Then, we use this value to obtain the corresponding intrinsic bolometric luminosity
by assuming an intrinsic AGN spectral shape such as the one given by, e.g., \citealp{elvis94}. The results of the simulations are presented in Figure~\ref{cloudy_res}. The 
best-fitting value for $\log$U can be found between -2.9 and -2.6, with a mean value of -2.7.

As can be seen in Figure~\ref{cloudy_res}, the resulting fits are not very sensitive to the value of n$_H$. Indeed, models with the same U and $N_H$ but different values 
of n$_H$ yield very similar reduced $\chi^2$. This is critical in our calculation, as the required intrinsic AGN bolometric luminosity is proportional to U$\times$$n_H$, 
as can be seen in equation \ref{eq:lum}. For the $n_H$ calculation we assume a fully-ionized gas, as was done for the Voorwerpjes \citep{keel12a}. Hence, the atomic 
hydrogen column density is the same as the electron density, $n_H$=$n_e$, which can be estimated from the ratio of the two [SII] lines. Furthermore, we can 
neglect the contribution from molecular hydrogen, given that no molecular gas was detected at the position of the southern ionized cloud. From the observed values 
of the [SII] 6716/6731 ratio we estimate an allowed average $n_H$ range in the southern emission region of 10-100 atoms cm$^{-3}$. In turn, this implies 
bolometric AGN luminosities in the range of 2.7$\times$10$^{45}$ erg s$^{-1}$ to 1.8$\times$10$^{46}$ erg s$^{-1}$. We
compare this value with the current AGN luminosity, as observed in X-rays. From {\it Chandra} observations, \citet{bianchi08} reported a 2-10 keV luminosity of 
1.5$\times$10$^{43}$ erg s$^{-1}$. Using their reported $N_H$ value of 7.1$\times$10$^{23}$~atoms cm$^{-2}$, we can estimate an intrinsic 2-10 keV luminosity 
of 9.9$\times$10$^{43}$ erg s$^{-1}$. Assuming a bolometric correction from 2-10 keV of 10 \citep[e.g.,][]{marconi04}, this implies that the intrinsic AGN luminosity 
was a factor of $\sim$3-20 times higher $\sim$40,000 years ago, when the light that photo-ionized the southern region was emitted. This is similar to the sample 
of Hanny's Voorwerp-like objects studied by \citet{keel17}.

Short term (up to tens of years) AGN variability by factors of $\sim$5 has been observed in the hard X-ray luminosity before 
\citep[e.g.,][and references therein]{ulrich97}. Hence, our results imply that the AGN activity in Mrk\,463E could have changed by up to an order of magnitude for the 
last $\sim$40,000. This is consistent with so-called ``standard'' variability and thus consistent with the AGN lifetime of $\sim$100,000 years 
suggested by \citet{schawinski15}.

\subsection{Feeding the SMBHs}

The {\it Chandra} X-ray data show that the SMBH in the Mrk\,463E nucleus is accreting at a $\sim$5$\times$ higher rate, and is more obscured than the one in
Mrk\,463W. In turn, most of the larger scale structures detected by the VLT/MUSE observations both in continuum and emission lines, appear to be 
associated preferentially with the Mrk\,463E nucleus. Similarly, both the southern emission line region and the outflowing cloud in the north appear
to be well aligned with the apparent biconical gradient, which in turn is also linked to the Mrk\,463E AGN. This could be interpreted as the AGN in Mrk\,463E being
older and more evolved than the one in Mrk\,463W, which is then most likely just starting its activity. Additional evidence pointing in that direction was presented 
on the previous section, where a lifetime greater than 40,000 years was estimated for the Mrk\,463E AGN.

Based on the ALMA $^{12}$CO(2-1) observations presented on Section~\ref{sec:morph_co} we see that both nuclei contain significant molecular gas reservoirs, 
$\sim$10$^9$M$_\sun$ around Mrk\,463E and $\sim$5$\times$10$^8$M$\sun$ surrounding Mrk\,463W. We further find $\sim$3$\times$10$^7$ M$\sun$ in 
molecular gas in clouds in between the two nuclei. This amount of material, albeit small, could be readily available to feed both SMBHs in the near future. Indeed, as 
presented in Section~\ref{sec:kine_co}, we estimate that the gas infalling rate onto the Mrk\,463E nucleus is $>$3 orders of magnitude larger than the SMBH accretion 
rate inferred from the X-ray luminosity. Hence, we can conclude that both nuclei have the potential fuel to feed quasar-like luminosity systems and increase their 
SMBH masses significantly, by adding $>$10$^8$~M$\sun$. However, and as presented in \S4.1, the outflow rate of the ionized gas is similar to the infall rate of the 
molecular gas onto the central regions of the system. Therefore, while the accretion rate onto the SMBHs is relatively small, there appears to be a rough balance
between the material infalling (as molecular gas) and outflowing (as ionized gas). This outflowing material does not appear to affect however the star formation
processes, as it is highly anisotropic and not aligned with the stellar disks.

\subsection{Possible Evolution}

Currently the two nuclei in Mrk\,463 are both active, perhaps at a relatively low level, heavily obscured at $N_H$$\sim$10$^{23}$ atoms cm$^{-2}$ and
separated by $\sim$4 kpc. As the merger continues, it is expected that the two nuclei will get closer over timescales of 10$^8$ years. The molecular gas that 
is now surrounding each nucleus will likely tend to concentrate such that a fraction of it will likely be available for accretion onto the SMBHs, while at the same time 
increasing the nuclear obscuration. Depending on the exact and yet unknown dynamics of the nuclei, the system may pass through a stage analogous to that which the 
nearby dual AGN NGC6240 is experiencing now, in which both nuclei are separated by $\sim$1 kpc and are heavily obscured \citep{komossa03}, while most of the 
gas is concentrated in between the two nuclei \citep[][Privon et al. in prep.]{tacconi99,iono07}. Alternatively, the gas could remain concentrated on each nucleus as they 
approach each other, thus creating a configuration similar to what it is observed now in the central region of Arp 220 \citep[e.g.,][and references therein]{manohar17}. While 
the amount of ongoing SMBH accretion in Arp 220 is still uncertain \citep[e.g.,][]{barcos-munoz15}, {\it Chandra} \citep{iwasawa05} and NuSTAR observations 
\citep{teng15} suggest that there is at least one heavily obscured, Compton thick, AGN associated with the system, on the western nucleus. In any case, following 
the scenario suggested by \citet{treister10}, we can expect that after $\sim$1-2$\times$10$^8$ years the collapsed nuclear region will start to become unobscured 
due to the effects of radiation pressure, thus revealing an unobscured quasar, perhaps similar to the one observed in Mrk 231, with strong ionized winds and 
ultrafast outflows \citep{feruglio15}, as those already observed in this system, associated with the Mrk 463E nucleus.

We note that the case of Mrk\,463 highlights the difficulty of using the instantaneous AGN luminosity in placing objects within evolutionary sequences. At least one
nucleus show evidence for changing luminosities on timescales of $10^{4-5}$ years, which is significantly shorter than those expected for dynamical evolution of the 
system \citep[e.g.,][]{lamassa15,parker16,gezari17}. This suggests it is likely problematic to use AGN luminosity even as a broad indicator of evolutionary stage, as at 
any point one or both nuclei can rapidly rise or drop in luminosity. This further highlights the need for multi-wavelength spatially-resolved spectroscopy over large 
spatial scales, as presented here, in order to fully understand the evolutionary stage of these complex merging galaxies.

\section{Conclusions}

We have presented optical and near-IR seeing-limited IFU data for the nearby dual AGN Mrk\,463, complemented by relatively high resolution band 6
ALMA data covering both the $^{12}$CO(2-1) transition and the surrounding $\sim$220 GHz continuum. The multiwavelength spatially-resolved spectroscopy 
reveal a very complex system in which the gas morphology is highly distorted, both due to the ongoing major merger and the SMBH accretion activity. Optical emission 
lines such as [O{\sc iii}] and H$\beta$ reveal the presence of AGN-ionized emission line regions extending beyond 10 kpc away from the nuclear regions, while the H$
\alpha$ and Pa$\alpha$ emission lines trace star forming regions up to even larger distances. The HeII 4686 line traces the influence of the AGN both in the nuclei and 
the extended regions. We find in general relatively low optical extinction in this system,  $A_{H\alpha}$$<$1, although moderate values are found toward 
specific isolated regions. Kinematically, the [O{\sc iii}] and H$\beta$ maps show evidence for a clear biconical outflow in the central region of the Mrk\,463E galaxy
 in which we infer mass outflow rates of $\sim$512 $M_\odot$/s. We further detected an outflowing region at $>$600 km s$^{-1}$, about 15 kpc to the north of the 
Mrk\,463E nucleus. Velocity gradients in these emission lines could also be detected in the tidal tails, most likely associated with the dynamics of the galaxy merger. From 
an energetics analysis of the southern emission line region, which according to the observed line ratios is consistent with being photo-ionized by the AGN emission, we 
conclude that the AGN luminosity, and thus SMBH accretion rate, on the Mrk\,463E nucleus changed by a factor 3-20 in the last 40,000 years, much like the handful 
of known Hanny's Voorwerp-like objects. The structure of the stellar kinematics reveals the presence of a strong velocity gradient to the west of the Mrk\,463W nucleus. 
While this gradient can be fitted by a Keplerian disk, it can also be associated to the tidal plumes originating from the ongoing major merger, thus suggesting that the 
system is undergoing a second or later close pass of the nuclei. Indeed, the rather smooth optical continuum spatial profile of that region, and the spatial separation 
of $\sim$3-4~kpc between the presumed stellar disk and the Mrk\,463W nucleus strongly supports the latter interpretation.

From the ALMA data we can see significant molecular gas reservoirs, $\sim$10$^9$M$_\sun$ in Mrk\,463E and $\sim$5$\times$10$^8$M$_\sun$ on the western 
nucleus, surrounding each nuclei, closer than $\sim$1 kpc but not directly on the AGN locations. The molecular gas around each nuclei present significant velocity 
gradients, thus suggesting a ring-like structure. While it is expected that a significant fraction of this molecular gas should be available for the SMBH to accrete, by 
comparing the molecular gas directly overlapping with the AGN with the accretion rate inferred from the AGN luminosity we conclude that only a very small fraction, 
$<$0.01\%, of the surrounding material is actively feeding the SMBH. Considering the observed properties of the gas in this system, we conclude that
in this system molecular gas is infalling onto the central regions at a rate of $\sim$100s M$_\odot$/yr. Of this, a relatively small fraction, $\sim$0.0017 M$_\odot$/yr,
is accreted onto the SMBH. This small amount of accreted material is able to generate an outflow of ionized gas that pushes $\sim$500 M$_\odot$/yr away from the 
nucleus. Hence, we can consider this major galaxy merger as a machine that receives large amounts of molecular gas and in return expels that material in the form 
of collimated ionized winds, being fueled by only a very small fraction, 10$^{-5}$, of that material being accreted by the SMBH.

We can speculate that in the future, as the two nuclei get closer to each other, the surrounding molecular gas will concentrate on the coalescent nucleus increasing both 
the SMBH accretion and nuclear obscuration. After $\sim$100 million years, the radiation pressure should be strong enough to evaporate most of the surrounding material 
revealing an unobscured high-luminosity AGN, i.e., quasars, characterized by strong winds and outflows as those already observed in nearby galaxies such as Mrk 231. 
High resolution dynamical modeling of this system would allow us to better predict the future of gas-rich major galaxy mergers.

While these multiwavelength observations have been critical to increase our understanding of the gas in the dual AGN Mrk\,463, it is clear that higher resolution data are 
necessary in order to understand for example the behavior of the gas actively feeding the SMBHs now and its connection to the merger dynamics. It is possible now to 
obtain these data thanks to the long baseline, $>$10 km, modes available with ALMA, which would yield spatial resolutions of tens of parsec  for the molecular gas 
observations. Similarly, observations of higher-J CO transitions would allow to measure physical parameters such as temperature and study the denser material, which 
might be located closer to the SMBHs. 

\acknowledgments

We thank the anonymous referee for a very positive and constructive report that helped improving this paper. We acknowledge support from: CONICYT-Chile 
grants Basal-CATA PFB-06/2007 (FEB, ET,CR), FONDECYT Regular 1141218 (FEB,CR) and 1160999 (ET), 
FONDECYT Postdoctorado 3150361 (GCP), EMBIGGEN Anillo ACT1101 (FEB, ET, GCP,CR); the Ministry of Economy, Development, and Tourism's Millennium 
Science Initiative through grant IC120009, awarded to The Millennium Institute of Astrophysics, MAS (FEB); Swiss National Science Foundation Grants 
PP00P2 138979 and PP00P2 166159 (KS, LS), and the China-CONICYT fund (CR). VU acknowledges funding support from the University of California 
Chancellor's Postdoctoral Fellowship.This paper makes use of the following ALMA data: 
ADS/JAO.ALMA\#2013.1.00525.S. ALMA is a partnership of ESO (representing its member states), NSF (USA) and NINS (Japan), together with 
NRC (Canada), NSC and ASIAA (Taiwan), and KASI (Republic of Korea), in cooperation with the Republic of Chile. The Joint ALMA Observatory is 
operated by ESO, AUI/NRAO and NAOJ. The National Radio Astronomy Observatory is a facility of the National Science Foundation operated under 
cooperative agreement by Associated Universities, Inc. This work is partly sponsored by the Chinese Academy of Sciences (CAS), through a grant to the CAS 
South America Center for Astronomy (CASSACA) in Santiago, Chile.



\vspace{5mm}
\facilities{VLT:Yepun (MUSE and SINFONI), ALMA, CXO}

\software{Fluxer, IRAF, CLOUDY, IDL, CASA, FUS, numpy, matplotlib, pyspeckit}

\end{document}